\documentclass[12pt]{iopart}
\usepackage{amssymb}
\usepackage{iopams}
\usepackage{cite}
\usepackage{graphicx}
\usepackage{xcolor}
\newcommand{\ds}{\displaystyle}

\begin{document}

\title{Local coordinates and motion of a test particle in the McVittie spacetime}

\author{Vishal Jayswal\footnote{Corresponding author: Vishal Jayswal, vjayswal@missouri.edu}, Sergei M. Kopeikin}

\address{Department of Physics \& Astronomy, University of Missouri-Columbia, Columbia, MO 65211, USA}
\ead{vjayswal@missouri.edu}
\vspace{10pt}
\begin{indented}
\item[]February 2025
\end{indented}

\begin{abstract}
	We consider the orbital motion of a test particle in the gravitational field of a massive body (that might be a black hole) with mass $m$ placed on the expanding cosmological manifold described by the McVittie metric. We introduce the local coordinates attached to the massive body to eliminate nonphysical, coordinates-dependent effects associated with Hubble expansion. The resultant equation of motion of the test particle are analyzed by the method of osculating elements with application of time-averaging technique. We demonstrate that the orbit of the test particle is not subject to the cosmological expansion up to the terms of the second order in the Hubble parameter. However, the cosmological expansion causes the precession of the orbit of the test particle with time and changes the frequency of the mean orbital motion. We show that the direction of motion of the orbital precession depends on the Hubble parameter as well as the deceleration parameter of the universe. We give numeric estimates for the rate of the orbital precession with respect to time due to the cosmological expansion in case of several astrophysical systems.
\end{abstract}
\vspace{2pc}
\section{Introduction}\label{sc2}
The study of gravitational effects produced of a black hole embedded into cosmological spacetime is an important area of research for understanding the evolution of self-gravitating isolated astronomical systems in the presence of Hubble's expansion. There are two important aspects of this problem: the study of the intrinsic coupling of the black hole mass with the cosmological scale factor $\mathfrak{a}\equiv\mathfrak{a}(t)$ \cite{mcvittie1933, Farrah_2023, GV2023}, and research on the evolution of orbit of a test particle gravitationally bound to a black hole. An adequate and comprehensive study of these aspects demands a rigorous mathematical approach admitting that spacetime of an isolated astronomical body (black hole) is not Minkowskian at infinity but matches asymptotically with the cosmological manifold of FLRW universe.

Several metrics, which are exact solutions of Einstein's equations, have been constructed to account for this particular property of spacetime of the isolated black hole. The first metric was the Schwarzschild-de Sitter metric discovered by Kottler \cite{1918AnP...361..401K}
\begin{eqnarray}\label{eqVIIIa1}
\hspace{-1.5cm}	{\rmd} s^2 =  -\left(1- \ds \frac{2m}{R}-H^2 R^2 \right) {\rmd} {T}^2 +  \left(1- \ds \frac{2m}{R}-H^2 R^2 \right)^{-1}{\rmd} R^2+ R^2 \;{\rmd} \Omega^2\,, 
\end{eqnarray}
where $H$ is a constant Hubble parameter usually associated with the Lambda parameter of the de-Sitter space $\left(H^2 =\frac{1}{3} \Lambda\right)$, ${\rmd} \Omega^2 = {\rmd} \theta^2 + \sin^2\theta ~{\rmd} \phi^2$, and $R$ is the radial Kottler coordinate.

McVitttie \cite{mcvittie1933} significantly extended the approach offered by the Kottler metric to study the gravitational field of a black hole embedded into cosmological FLRW spacetime. In doing so, he found an exact solution of Einstein's field equations which became known as McVittie's metric. The metric is built in the global isotropic coordinates, $x^{\alpha} \equiv (t, r, \theta, \phi$) covering the entire FLRW manifold and it is given by
\begin{eqnarray}\label{eq50}
	{\rmd} s^2 = -{\rm e}^{\zeta}\;{\rmd} t^2 + {\rm e}^{\nu}\;\left[{\rmd} r^2 + r^2\;\left({\rmd} \theta^2 + \sin^2\theta\; {\rmd} \phi^2\right)\right]\,,
\end{eqnarray} 
where, 
\begin{eqnarray}\label{eq151b}
	{\zeta} = - 2\,{\rm\ln} \left[\ds \frac{1+\ds \frac{\mu(t)}{2r}\sqrt{1+\ds\frac{kr^2}{4}}}{1-\ds \frac{\mu(t)}{2r}\sqrt{1+\ds\frac{kr^2}{4}}}\right]\,,
\end{eqnarray}	
\begin{eqnarray}\label{eq151c}
	{\nu} = 2 {\rm\ln}\, \mathfrak{a}(t)  + 4\, {\rm\ln}\,{\left[ \ds \frac{1}{\sqrt{1+\ds\frac{kr^2}{4}}}+\ds \frac{\mu(t)}{2r}\right]}\,.
\end{eqnarray}
In equations \eref{eq151b} and \eref{eq151c} $\mathfrak{a}(t)$ represents a cosmological scale factor, $\mu(t)={m}/{\mathfrak{a}(t)}$ is a mass function expressed in terms of a constant mass $m$ of an isolated astronomical body, and the cosmological scale factor $\mathfrak{a}(t)$. 

It is also worth mentioning two other exact solutions for a point-like mass in a cosmological background. The first solution, found by Nandra \etal. \cite{NLH2012}, replicates McVittie's metric for a flat universe, but its physical meaning for open and closed universes remains better understood. The second solution, obtained by Faraoni \etal. \cite{Faraoni_2014}, incorporates both the mass $m$ and the electric charge $Q$ of a black hole, extending the McVittie spacetime for a flat universe. This provides valuable theoretical insights into the behavior of charged black holes in an expanding cosmological background. Nonetheless, real astrophysical systems are electrically neutral. Therefore, we consider only a one-parameter solution depending solely on the mass $m$ of the central body placed in the cosmological background manifold.

The Schwarzschild and FLRW metric are two limiting cases of McVittie's metric. Indeed, the metric given by equation \eref{eq50} reduces to the Schwarzschild metric in the isotropic coordinates in the limit $k \rightarrow 0$ and $\mathfrak{a}(t) \rightarrow 1$ \cite{1973grav.M},
\begin{eqnarray}\label{eqXXIII}
	{\rmd} s^2 =  -\left(\frac{\ds 1-\frac{m}{2r}}{1+ \ds \frac{m}{2r}}\right)^2 \rmd t^2 +  \left(1+\frac{m}{2r}\right)^4\left({\rmd} r^2 + r^2 \;{\rmd} \Omega^2\right)\,.
\end{eqnarray}
The FLRW metric can be obtained from McVittie's metric by imposing the limit
$m \rightarrow 0$. In this case, equation \eref{eq50} reduces to the FLRW metric in isotropic coordinates,
\begin{equation}\label{eqXXV}
	{\rmd} s^2 =  - {\rmd} t^2 + \frac{\mathfrak{a}^2(t)}{\ds\left(1+\frac{kr^2}{4}\right)^{2}}\;\left({\rmd} r^2 + r^2 \;{\rmd} \Omega^2\right)\,.
\end{equation}
McVittie's metric depends on the mass function
\begin{eqnarray}\label{eq6aee}
	\mu(t) =\ds \frac{m}{\mathfrak{a}(t)}\,,
\end{eqnarray}
where mass $m$ is constant, but is intrinsically coupled to the cosmological scale factor. Some researchers \cite {Croker_2021} have suggested that the mass of the central body couples with the cosmological scale factor in a more general way,
\begin{eqnarray}\label{eq6ae1}
	\mu(t)=\ds \frac{m}{\mathfrak{a}^n(t)}\,,
\end{eqnarray}
where $n$ is the coupling index. Such proposal might have interesting astrophysical consequences for cosmological evolution of black holes \cite {Farrah_2023, Lacy_2024}. Unfortunately, exact solutions of Einstein's field equations corresponding to the case $n\ne1$ have not yet been found. So far, all known exacts cosmological solutions of Einstein's field equations in the presence of a point-like mass (McVittie, Schwarzschild-de Sitter, etc) suggest the coupling index $n=1$ in equation \eref{eq6ae1}, which means the cosmological expansion does not affect the mass of the central body \cite{GV2023} since all these solutions are reduced to the metric, equation \eref{eqXXIII}, of the Schwarzschild black hole locally. 
		
Besides the exact solutions, there exists a powerful field-theoretical approach to finding an approximate solution to Einstein's equations that describes the gravitational field of an isolated astronomical system embedded in a cosmological background \cite{Kopeikin_2015FTPK, Kopeikin_2014AnPhy}. Some progress in finding such a solution has been made in \cite{Kopeikin_2012PhRvD, Kopeikin_2016PhRvD}. Unfortunately, this solution has not yet been developed to include a sufficient number of post-Newtonian terms in the expansion with respect to the Hubble parameter $H$, and is therefore deficient for the purposes of the present paper.

In this paper, we adopt McVittie's metric to consider the evolution of orbit of a test particle around the central body with mass $m$. The original form of McVittie's metric \cite{mcvittie1933} has been derived in the global cosmological coordinates which are not suitable for interpretations of observations conducted by a local observer because they have been designed mostly for exploration of the global aspects of the problem. Physical interpretation of observations of the orbital behavior of the test particle is conducted by an observer associated with the central massive body straightforward in the local coordinates of the observers, which we shall denote as $X^\alpha = (T,R,\theta,\phi)$. 

Transformation to the local coordinates, $T=T(t,r), R=R(t,r)$ can be found with several techniques. The most straightforward way to introduce the local radial coordinate on the background FLRW cosmological spacetime would be 
\begin{eqnarray}\label{eq3a1}
	R = \mathfrak{a}(t) r \,,
\end{eqnarray}
which is often used for the definition of the proper distance in cosmology \cite{weinbergbook1972}. However, this equation is valid only in a linear approximation with respect to the radial coordinate. More complicated transformation was proposed by Klioner and Soffel \cite{2005ESASP.576..305K} who extended equation \eref{eq3a1} to higher-order approximations. Transformation of McVittie's metric to the local coordinates on the basis of equation \eref{eq3a1} is insufficient as it does not take into account the effect of the central mass $m$ on the transformation. Robertson \cite{doi:10.1080/14786440508564528} has found a transformation of the Kottler metric \eref{eqVIIIa1} for such case,
\begin{eqnarray}\label{eq3a}
R =\left(1+ \ds \frac{\mu(t)}{2 r}\right)^2 \mathfrak{a}(t) r \,,
\end{eqnarray}
thus, demonstrating that the Kottler metric is a particular case of McVittie's metric for $k=0$, $\mathfrak{a}(t)= {\rm exp} (H t)$ with $H= {\rm constant}$. Later on, Carrera and Giulini \cite{carrera2010} used the transformation \eref{eq3a} to study the motion of the test particle in McVittie's spacetime of the spatially-flat FLRW universe $k=0$. It should be emphasized, however, that transformation \eref{eq3a} is not applicable to the case of open ($k=-1$) and closed universe ($k=+1$). Generalization of the transformation \eref{eq3a} to the cases of open and closed FLRW universes requires more elaborated mathematical technique. The transformation \eref{eq3a} should be extended in the case of more advanced solutions to account for additional parameters, such as the electric charge $Q$ of a black hole \cite{Faraoni_2014}.

In particular, Nandra \etal. \cite{NLH2012} used the tetrad formalism to introduce the transformation of McVittie's metric to the local coordinates in all possible models ($k= -1, 0, +1$) of the cosmological FLRW spacetime perturbed by the point-like mass $m$. Tetrad basis is formed by assigning to each point $x^\mu$ of space-time a set of four unit vectors 
\begin{eqnarray}\label{eq3}
	\mathbf{\hat{e}}_{\alpha} = \{\mathbf{\hat{e}}_{0}, \mathbf{\hat{e}}_{1}, \mathbf{\hat{e}}_{2}, \mathbf{\hat{e}}_{3}\}\,,
\end{eqnarray}
which form a locally inertial frame (or local Lorentz frame) at each point of the spacetime manifold. Normalization of the basis vectors is defined in terms of the Minkoswki metric  $\eta_{\alpha \beta} = {\rm{diag}}\:(-1,+1,+1,+1)$ such that
\begin{eqnarray}\label{eq4}
	\mathbf{\hat{e}}_\alpha \cdot \mathbf{\hat{e}}_\beta =\eta_{\alpha \beta}\,,
\end{eqnarray}
where the dot between two vectors denotes a scalar product. The relationship between tetrad basis $\mathbf{\hat{e}}_\mu$  and coordinate basis $\mathbf{e}_\nu$ is defined as a linear transformation
\begin{eqnarray}\label{eq6}
		\mathbf{\hat{e}}_\mu  = \Lambda_\mu{}^\nu~\mathbf{e}_\nu\,, \qquad
		\mathbf{e}_\mu  = K^\mu{}_\nu~\mathbf{\hat{e}}_\mu\,,\label{eq6ac}
\end{eqnarray}
where $\Lambda$ restore the indices of the matrices and $K$ are the matrices of the direct and inverse transformation which are complementary to one another in the sense that 
\begin{eqnarray}\label{eq6a}
	\Lambda_\mu{}^\alpha~K^\mu{}_\alpha = \delta^\alpha{}_\beta \,.
\end{eqnarray}
In case of a spherically-symmetric and time-dependent gravitational field, it is instructive to define the components of the inverse matrix $K^\mu{}_\alpha$ by scalar functions $f_1 \equiv f_1(t,R)$, $g_1 \equiv g_1(t,R)$, and $g_2 \equiv g_2(t,R)$ as follows \cite{lasenbybook2007}: 
\begin{eqnarray}\label{eq7e}
\hspace{-2.3cm} K^0{}_0 = \frac{1}{f_1},\qquad K^1{}_0 = - \frac{g_2}{f_1 g_1}, \qquad K^1{}_1 = \frac{1}{g_1}, \qquad  K^2{}_2 = R, \qquad K^3{}_3 = R\sin\theta\,.
\end{eqnarray}
The metric tensor $g_{\mu\nu}$ is expressed in terms of the components of the inverse matrix $K^\mu{}_\alpha$ by making use of equation \eref{eq6ac} which yields 
\begin{eqnarray}\label{eq7b}
	g_{\mu\nu}=\mathbf{e}_\mu \cdot \mathbf{e}_\nu
	 =\eta_{\alpha \beta}~K^\alpha{}_\mu~K^\beta{}_\nu\,,
\end{eqnarray}
or, more explicitly,  
\begin{equation}\label{eq162a}
	{\rm d} s^2 = - \left(\ds \frac{g_1^2 - g_2^2}{f_1^2g_1^2}\right)~{\rm d} t^2 - \ds \frac{2g_2}{f_1g_1^2}~{\rm d }t\,{\rm d} R+ \ds \frac{1}{g_1^2}~{\rmd }R^2 + R^2~{\rm d}\Omega^2 \,.
\end{equation}
Doran and Lasenby \cite{lasenbybook2007} used the tetrad formalism to formulate Einstein's field equations for the case of a massive body embedded to the FLRW universe, as a linear system of partial differential equations in the auxiliary local coordinates $(t,R, \theta, \phi)$ for functions $f_1 \equiv f_1(t,R)$, $g_1 \equiv g_1(t,R)$, and $g_2 \equiv g_2(t,R)$ entering equation \eref{eq162a}. The reader should notice that the time coordinate in the Doran-Lasenby approach remains the same as the Hubble time $t$. Nandra \etal. \cite{NLH2012} solved these equations for the general case of FLRW background cosmological manifold perturbed by a massive point-like particle for different values of spatial curvature $k=(-1, 0, +1)$. They have found their solution equal to McVittie's metric for $k=0$. However, Lasenby's metric \eref{eq162a} for closed $(k=+1)$ and open $(k=-1)$ universes differs from McVittie's metric. Nonetheless, the Lasenby approach was instrumental for better understanding of the nature of the transformation from the global to local coordinates on cosmological manifold perturbed by a massive body.

Nandra \etal. \cite{NLH2012} used $f_1$, $g_1$, and $g_2$ components of their metric to derive geodesic equation of a test particle moving around the central massive body in an expanding universe and found a general-relativistic expression for the force required to hold a test particle at rest relative to the central mass. Carrera and Giulini \cite{carrera2010} have conducted a similar study in the context of McVittie's spacetime but only in case of the FLRW universe with $k=0$ which limits their range of applicability.

In this section, we discussed McVittie's metric \eref{eq50} in the global coordinates and justified the need for the local coordinates affiliated with the central black hole. In next sections we implement the technique of the tetrad formalism to the McVittie metric to build the local coordinates of physical observer and to study behavior of closed elliptical orbits of test particles around the central black hole in the expanding FLRW for all the three types of spatial curvature: $k=(-1,0,+1)$. More specifically, we employ in section~\ref{scPF-McV} the Lagrange Inversion Theorem to transform the McVittie metric \eref{eq50} from the global coordinates $x^\alpha \equiv (t,r, \theta,\phi)$ to the local coordinates $X^\alpha\equiv(T,R,\theta,\phi)$, where the time $T$ is the coordinate time of the inertial frame of the local observer. Because spacetime is spherically symmetric the angular coordinates $(\theta, \phi)$ remain the same. \Sref{sc4} derives the post-Newtonian equations of motion of a test particle in the post-Friedmannian approximation of the FLRW universe including terms that are quadratic with respect to the Hubble parameter $H$. These equations are integrated in sections \ref{scpN}-\ref{sck} by the method of osculating elements and time-averaging technique \cite{brumbergbook1991}.
We discuss the result of the integrations in \sref{scCon} by providing numerical estimates of the secular change of the orbital osculating elements for the orbits of stars moving around the supermassive black hole at the Milky Way galactic center, as well as for a binary star Sirius.

The paper includes three appendices. Appendix A demonstrates that Kottler's metric (Schwarzschild-de Sitter metric) \cite{doi:10.1080/14786440508564528} corresponds to McVittie's metric in local coordinates for a flat universe ($k=0$). Appendix B provides detailed technical information on the application of the averaging technique. Appendix C discusses the exact asymptotic solution of the equations governing the evolution of the orbital elements $a$ and $e$, and shows that this solution aligns with the results obtained through the averaging method.

In this paper we use geometrical system of units with $ G = c = 1$, where $G$ is the universal gravitational constant and $c$ is the speed of light in free space. The other notations are as follows:

\noindent -- Latin indices $i, j, k,$ label spatial coordinates and take values 1, 2, 3.

\noindent -- Greek indices $\alpha,\beta,\gamma,$ label spacetime coordinates and take values 0, 1, 2, 3.

\noindent -- Repeated indices indicate the Einstein summation rule.

\noindent -- $\eta_{\alpha \beta} = \rm{diag}\;\left(-1, +1, +1, +1\right) $ is the Minkowski metric and $g_{\alpha \beta}$ is the space-time metric.

\noindent -- A dot over any quantity denotes a partial time derivative, and a prime over does a partial radial derivative.

\noindent -- $\mathfrak{a}(t)$ is the scale factor of the cosmological model.

\noindent -- $ H(t)= {\dot{\mathfrak{a}}}/{\mathfrak{a}}$ is the Hubble parameter with approximate numerical value at the present epoch $H_0=71\; {\rm km\:s^{-1}\:Mpc^{-1}} = 2.3 \times 10^{-18}\; {\rm s}^{-1}$.

\noindent -- $\mathfrak{q}$ is dimensionless deceleration parameter present in the equation $\dot{H} = -H^2 \left(1+\mathfrak{q}\right)$.

\noindent -- $k$ is the constant curvature of space taking one of three values: $-1,\; 0,\;+1$.

\noindent --$M$ and $m ={GM}/{c^2}$ are the constant mass of the central body in SI and geometrized units respectively. 

A generalized hypergeometric function that appears in \sref{sc3} is defined by 
\begin{eqnarray}
	 {}_m  F_n\left[\alpha_1,\alpha_2,...\alpha_m;\beta_1,\beta_2,...\beta_n;z\right]=\sum_{r=0}^{\infty}\ds \frac{\left(\alpha_1\right)_r\left(\alpha_2\right)_r...\left(\alpha_m\right)_r}{\left(\beta_1\right)_r\left(\beta_2\right)_r...\left(\beta_n\right)_r} \ds \frac{z^r}{r{!}}\,, \label{eq179aa}
\end{eqnarray}
where $\left(\alpha\right)_n$ is the Pochhammer symbol 
\begin{eqnarray}\label{eq179bb}
		\left(\alpha\right)_n = \ds\frac{\Gamma\left(\alpha+n\right)}{\Gamma\left(\alpha\right)}=\alpha \left(\alpha+1\right)...\left(\alpha+n-1\right)\,,
\end{eqnarray}
with $\left(\alpha\right)_0=1$. Gamma function $\Gamma\left(n\right)$ is expressed as
\begin{eqnarray}\label{eq179cc}
	\Gamma\left(n\right)= \left(n-1\right)!=1\cdot2\cdot3\cdot...\cdot\left(n-1\right)\,.
\end{eqnarray}
\section{McVittie's metric in the local coordinates $X^\alpha = (T, R, \theta, \phi)$}\label{sc3}
Isotropic coordinates $x^\alpha = (t, r, \theta, \phi)$ match smoothly with the comoving coordinates of the Hubble observers of the background FLRW universe. These coordinates cover the entire cosmological manifold and are adequate for description of the global evolution of the universe. However, the metric of FLRW universe expressed in comoving coordinates does not satisfy the principle of equivalence locally. Therefore, description of motion of test particles in the vicinity of central mass $m$ in global isotropic coordinates $x^\alpha = (t, r, \theta, \phi)$ has many coordinate-dependent effects that are not physical.

To interpret local physical experiments, one needs to transform McVittie's metric to the local coordinates which are consistent with Einstein's equivalence principle (EEP). We shall denote such local coordinates as $X^\alpha = (T, R, \theta, \phi)$. 
These coordinates will be defined in such a way that the McVittie metric is reduced to the Minkowski metric at the origin of the local coordinate system in the limit $m \rightarrow 0$. 

We build these local coordinates in two steps:
\begin{enumerate}
\item[1)] The radial coordinate $r$ is transformed to $R=R(r, t)$ while keeping the time coordinate unchanged, $t = t$, in such a way that the McVittie metric \eref{eq50} is transformed to the form \eref{eq162a} proposed by Lasenby \cite{lasenbybook2007}.
\item[2)] The time coordinate $t$ is transformed to $T=T(t,R)$ while keeping the new radial coordinate unchanged, $R=R$. This transformation is determined by the condition that the off-diagonal term of the metric \eref{eq162a} vanishes.
\end{enumerate}
\subsection{Transformation of the radial coordinate}\label{sbscR}
The transformation of the global radial coordinate $r$ to the local radial coordinate $R$ is given by equation 
\begin{eqnarray}
	R = r\; {\rme}^{\nu/2} =r \;\mathfrak{a}(t)\left[{\ds  \frac{1}{\sqrt{1+\ds\frac{kr^2}{4}}}+\ds \frac{\mu(t)}{2r}}\right]^2\,.\label{eq61a}
\end{eqnarray}
Transformation \eref{eq61a} brings McVittie's metric \eref{eq50} to the form \eref{eq162a} where functions $f_1 \equiv f_1(t,R)$, $g_1 \equiv g_1(t,R)$, and $g_2 \equiv g_2(t,R)$ are given by
\numparts 
\begin{eqnarray}
	f_1\equiv {\rme}^{-{\zeta/2}}\,,\label{eq158ee}\\
	g_1\equiv 1+\ds \frac{r}{2}{\ds \frac{\partial \nu}{\partial r}}\,,\label{eq158cc}\\
	g_2\equiv H R\,. \label{eq158dd}
\end{eqnarray}
\endnumparts
Here, functions $\zeta$ and $\nu$ are given by equations \eref{eq151b} and \eref{eq151c} respectively, the radial coordinate $r$ is understood as function of the local radial coordinate and time, $r=~r(T,R)$, and 
\begin{eqnarray}
	\ds \frac{\partial \nu}{\partial r}=\ds \frac{\sqrt{\mathfrak{a}(t)r}}{\sqrt{R}}\left[-\ds\frac{kr}{\left(1+\ds\frac{kr^2}{4}\right)^{3/2}}-\ds\frac{2\mu(t)}{r^2}\right]\,.\label{eq158b}
\end{eqnarray}
It is convenient to introduce a new function $w$ defined by
\begin{eqnarray}
	w =  \ds \frac{2}{\mu(t)}\frac{r}{\sqrt{1+\ds\frac{kr^2}{4}}} \,.\label{eq165a}
\end{eqnarray}
In terms of the function $w$ the local coordinate
\begin{eqnarray}
R=	\ds \frac{2m}{w}{\left(1+w\right)^2}{\sqrt{1-\ds \frac{k}{16}  \mu^2 w^2}}\,,\label{eq176}
\end{eqnarray}
and the tetrad component 
\begin{eqnarray}
	f_1 = \ds \frac{w+1}{w-1}\,.\label{eq167aa}
\end{eqnarray}
The tetrad component $g_1$ is not independent but relates to the function $f_1$ by
\begin{eqnarray}
	f_1g_1=\left[1- \ds \frac{k \mu^2}{8} \ds \frac{w^3}{w-1}\right]\,.\label{eq167}
\end{eqnarray}
\subsection{Transformation of the time coordinate}\label{sbscT}
We eliminate the off-diagonal component of the metric \eref{eq162a} with the help of the coordinate transformation, 
\begin{eqnarray}
	t = T - \chi(T,R)\,.\label{eq180a}
\end{eqnarray}
It brings the McVittie metric to the following form
\begin{eqnarray}
	\hspace{-1cm}	{\rmd} s^2 =  {G}_{TT}(T,R)\;{\rmd} T^2 + 2{G}_{TR}(T,R)\;{\rmd} T\,{\rmd} R + {G}_{RR}(T,R)\; {\rmd} R^2 + R^2\;{\rmd} \Omega^2\,, \label{eq182a}
\end{eqnarray}
where the components of the metric tensor are given by
\numparts
\begin{eqnarray}\label{eq183}
	{G}_{TT}(T,R) = - \left(\ds \frac{g_1^2 - g_2^2}{f_1^2g_1^2}\right) {t_T^2}\,,\label{eq183a}\\
	{G}_{TR}(T,R) = -\left[\left(\ds \frac{g_1^2 - g_2^2}{f_1^2g_1^2}\right)~{t_ R}+\ds \frac{g_2}{f_1g_1^2}\right]~{t_T}\,,\label{eq183b}\\
	{G}_{RR}(T,R) = - \left[\left(\ds \frac{g_1^2 - g_2^2}{f_1^2g_1^2}\right)~{t_R^2}+\left(\ds \frac{2g_2}{f_1g_1^2}\right)~{t_R}-\ds \frac{1}{g_1^2}\right]\,,\label{eq183c}
\end{eqnarray}
\endnumparts
where ${t_T} \equiv {\partial t}/{\partial T}$ and ${t_R} \equiv {\partial t}/{\partial R} $. The metric components $G_{TR}(T,R)$ can now be removed from the metric \eref{eq182a} by imposing a differential condition on function $\chi$ such that   
\begin{eqnarray}\label{eq184}
	& \ds \frac{\partial \chi}{\partial R} = \ds \frac{F_1G_2}{G_1^2 - G_2^2}\,,
\end{eqnarray}
where functions 
\numparts
\begin{eqnarray}
F_1\equiv F_1 (T,R) = f_1 [ t(T,R),R ]\,,\label{eq184a} \\
G_1\equiv G_1(T,R) = g_1 [ t(T,R),R ]\,,\label{eq184b} \\
G_2 \equiv G_2(T,R)= g_2 [ t(T,R),R ]\, ,\label{eq184c}
\end{eqnarray}
\endnumparts
and all constants of time like the scale factor and Hubble parameter are understood as functions of new time and radial coordinate, for example 
\begin{eqnarray}
	\mathfrak{a}(t) = \mathfrak{a}(T)  - \chi(T,R)\ds \frac{{\rmd} \mathfrak{a}(T)}{{\rmd} T} + ...\:,\label{eq183dd}\\
	H(t) = H(T)  - \chi(T,R) \ds \frac{{\rmd} H(T)}{{\rmd} T} + ... \:. \label{eq183ee}
\end{eqnarray}
\Eref{eq184} can be solved by integration 
\begin{eqnarray}
	\chi(T,R) = \int \ds \frac{F_1G_2}{G_1^2 - G_2^2} {\rmd} R + C(T)\,,\label{eq182b1}
\end{eqnarray}
where $C(T)$ is an arbitrary function of time that can be absorbed into the definition of new time coordinate $T$. The remaining non-vanishing components of the McVittie metric \eref{eq182a} are
\begin{eqnarray}\label{eq185a}
	{G}_{TT}(T,R) = - \left(\ds \frac{G_1^2 - G_2^2}{F_1^2G_1^2}\right) \left[1 - \ds \frac{\partial \chi(T,R)}{\partial T}\right]^2 \,,
\end{eqnarray}
\begin{eqnarray}
	{G}_{RR}(T,R)	 =\ds \frac{1}{G_1^2 - G_2^2}\,.\label{eq185}
\end{eqnarray}
The McVittie metric in the new local coordinates $(T,R)$ reads
\begin{eqnarray}
	{\rmd} s^2 = {G}_{TT}(T,R)\;{\rmd} T^2 + \left[{G}_{RR}(T,R)-1\right]\;{\rmd} R^2 + \delta_{ij}\;{\rmd} X^i\;{\rmd} X^j\,,\label{eq190}
\end{eqnarray}
where $X^1 = R\sin\theta\,\cos\phi,\; X^2 = R \sin\theta\,\sin\phi,\; X^3 = R \cos\theta$ are the Cartesian coordinates such that $R^2=\delta_{ij} X^i X^j$. The McVittie metric \eref{eq190} still represents an exact vacuum solution of the Einstein equations expressed in terms of the local coordinates~$(T,R)$. Unfortunately, its exact form is too complicated for analysis of motion of the test particle around the central black hole embedded in the expanding cosmological background. To render this analysis we shall expand the McVittie metric~\eref{eq190} with respect to several small parameters: $HR/c\ll1$, ${m}/{R}\ll1$, $m/{\mathfrak{a}\left(t\right)}\ll1$ up to the terms of the second-order.
\section{Post-Friedmannian approximation of McVittie's metric}\label{scPF-McV}
The coefficients of McVittie's metric are composite functions of the local coordinates through the coordinate transformations: $t = t(T,R)$ and $r = r(T,R)$ which are extremely difficult to work with directly. In order to circumvent this difficulty we expand the McVittie metric coefficients  in equation \eref{eq190} in the post-Friedmanian series with respect to the small parameter $HR/c$. To expand the McVittie metric \eref{eq190} with respect to the small parameters $HR/c$ we need to express the global radial coordinate $r$ in terms of the local coordinate $R$ explicitly. This can be achieved by solving  equation~\eref{eq176} for variable $w$ defined in equation \eref{eq165a}, which represents an algebraic polynomial equation of the sixth order with respect to the variable $w$. This equation can not be solved in radicals and, hence, we apply Lagrange's inversion theorem to find its roots.

In order to apply this theorem we have to convert equation \eref{eq176} to the following form
\begin{eqnarray}
	w=-1 - z\phi\left(w\right) \,,\label{eq176d}
\end{eqnarray}
where
\begin{eqnarray}
	z = \left({\ds \frac{2 R}{m}}\right)^{1/2}\qquad ,\qquad
	\phi\left(w\right) \equiv  w^{1/2}~\left(1-K w^2\right)^{-1/4}\,, \label{eq176c}
\end{eqnarray}
with $K \equiv \ds \frac{k\mu^2}{16}\ll1$. \Eref{eq176d} is solved by expanding function $w$ around the point $w=-1$ in the formal power series. According to Lagrange's inversion theorem, the solution of equation~\eref{eq176d} is given by
\begin{eqnarray}
	w =-1 + \sum_{n=1}^{\infty}\ds \frac{\left(-z\right)^n}{n!}  \lim_{w\to -1} \ds \frac{{\rmd}^{n-1}}{{\rmd} w^{n-1}}\left[\phi\left(w\right)\right]^n\,.\label{eq177a}
\end{eqnarray} 
In order to compute $(n-1)^{{\rm th}}$ derivative in \eref{eq177a} we integrate the factor $[\phi(w)]^n$ with respect to $w$ and substitute it back to \eref{eq177a}. It yields 
\begin{eqnarray}
	\hspace{-1.8cm} w =-1+\sum_{n=1}^{\infty} \ds \frac{\left(-z\right)^n}{n!}  \lim_{w\to -1} \ds \frac{{\rmd}^{n}}{{\rmd} w^{n}}\left[\ds \frac{2}{2+n}w^{1+\frac{n}{2}}~{}_2 F_1\left(\ds \frac{n}{4},\ds \frac{2+n}{4};\ds \frac{6+n}{4};K w^2\right)\right]\,,\label{eq177b}
\end{eqnarray}
where ${}_2F_1\left(\alpha_1,\alpha_2;\beta_1;z\right)$ is the Gauss hypergeometric function. \Eref{eq179aa} allows us to rewrite \eref{eq177b} in terms of the Pochhammer symbols
\begin{eqnarray}
	\hspace{-2.2cm}	w =-1+ \sum_{n=1}^{\infty} \frac{\left(-z\right)^n}{n!} \frac{2}{2+n}~\left[\sum_{p=0}^{\infty}{\ds \frac{\left(\ds \frac{n}{4}\right)_p\left(\ds \frac{2+n}{4}\right)_p}{\ds \left(\frac{6+n}{4}\right)_p}~\frac{K^p}{p!}}~ \lim_{w\to -1} \frac{{\rmd}^{n}}{{\rmd} w^{n}}\Bigl(w^{1+ \frac{n}{2}+2p}\Bigr)\right]\,,\label{eq177e}
\end{eqnarray}
where the limit is computed as follows
\begin{eqnarray}\label{eq177e1}
	\lim_{w\to -1} \ds \frac{{\rmd}^{n}}{{\rmd} w^{n}}w^{\ds\left(1+\ds \frac{n}{2}+2p\right)}=-\rmi^{-n} \left(2-\ds \frac{n}{2}+2 p\right)_n\,.
\end{eqnarray}
We use the dimidiation formula for Pochhammer symbols to obtain
\begin{eqnarray}
	\left(2-\ds \frac{n}{2}+2 p\right)_n =  \ds\frac{\Gamma\left(2+\ds\frac{n}{2}\right)}{\Gamma\left(2-{\ds\frac{n}{2}}\right)} \ds \frac{\left(1+\ds \frac{n}{4}\right)_p \left(\ds \frac{3}{2}+\ds \frac{n}{4}\right)_p}{\left(1-\ds \frac{n}{4}\right)_p \left(\ds \frac{3}{2}-\ds \frac{n}{4}\right)_p}\,.\label{eq177e12}
\end{eqnarray}
We substitute equations \eref{eq177e1} and \eref{eq177e12} back into equation \eref{eq177e} to obtain  
\begin{eqnarray}
	\hspace{-2.4cm}	w=-1 - \sum_{n=1}^{\infty}\ds \frac{(-z)^n}{n!} \ds \frac{2}{2+n}~\left[{\rmi}^{-n}\ds\frac{\Gamma\left(2+\ds\frac{n}{2}\right)}{\Gamma\left(2-{\ds\frac{n}{2}}\right)}\sum_{p=0}^{\infty}\ds{\ds\frac{\left(\ds\frac{2+n}{4}\right)_p\left(\ds\frac{4+n}{4}\right)_p\left(\ds\frac{n}{4}\right)_p}{\left(\ds\frac{4-n}{4}\right)_p\left(\ds\frac{6-n}{4}\right)_p}~\ds \frac{K^p}{p!}}\right].\label{eq177c}
\end{eqnarray}	
\Eref{eq177c} can be further simplified by expressing the Pochhammer symbols in equation~\eref{eq177c} in terms of the Gamma function. It gives solution of equation \eref{eq176d} in the final form 
\begin{eqnarray}
	w= -\sum_{n=0}^{\infty}\sum_{p=0}^{\infty}\ds \frac{\Gamma\left(\ds \frac{n}{4}+p\right)}{\Gamma\left(\ds \frac{n}{4}\right)} \ds \frac{\Gamma\left(1+2p+ \ds \frac{n}{2}\right)}{\Gamma\left(2+2p-\ds \frac{n}{2}\right)}\ds\frac{K^p}{n{}!\,p{}!}\ds \left(-\frac{2R}{m}\right)^{n/2}\,.\label{eq177f}
\end{eqnarray}
\Eref{eq177f} is used for writing down functions $f_1$ and $g_1$ defined by equations \eref{eq167aa}, and \eref{eq167} as explicit functions of the local coordinate $R$. We have 
\begin{eqnarray}
	f_1 =\ds \frac{1}{x} - \ds \frac{K \left(1+x\right)^3}{4 \left(1-x\right)\,x^3}+\mathcal{O} \left(K^2\right)\,,\label{eq177i}
\end{eqnarray}
\begin{eqnarray}
	g_1 = x+\ds \frac{K \left(1+x\right)^3 (1-5x)}{4 \left(1-x\right)^2 x}+ \mathcal{O}\left(K^2\right)\,,\label{eq182c}
\end{eqnarray}
where we have introduced a new variable
\begin{eqnarray}
 x \equiv \left({1-{2 m}/{R}}\right)^{1/2}\,,\label{eq182c1}
\end{eqnarray}
and dropped out the residual terms of the order $K^2$ because they are negligibly small. 

Equations \eref{eq177i} and \eref{eq182c} are used in order to calculate the time transformation function $\chi$ defined by \eref{eq182b1}, which can be simplified to 
\begin{eqnarray}
	\chi = H \int \ds \frac{F_1 }{G_1^2} R \;{\rmd} R + \mathcal{O}\left(H^2\right)\,,\label{eq182b}
\end{eqnarray}
by expanding denominator on right side of equation \eref{eq182b1} in power series with respect to parameter $HR/c\ll1$. Substituting $F_1$ and $G_1$ from equations \eref{eq204o4}, \eref{eq204o5} and integrating we get 
\begin{eqnarray}
		\chi=\ds \frac{1}{2}  H R^2 \left(1+\ds\frac{6 m}{R}\right)+k\ds \frac{H R^4}{4 \mathfrak{a}^2}\,,\label{eq186a}
\end{eqnarray}
where we have dropped all the terms of the quadratic order with respect to the small parameters. Partial derivative of the function $\chi(T,R)$ with respect to time coordinate~$T$ is given by 
\begin{eqnarray}
	\ds \frac{\partial \chi}{\partial T}=\ds \frac{1}{2}  \dot{H} R^2 \left(1+\frac{6m}{R}\right) -\ds \frac{\kappa}{4}\left(2 H^2-\dot{H}\right)R^4\,.\label{eq187a1}
\end{eqnarray} 
\Eref{eq186a} is substituted in equations \eref{eq183dd} and \eref{eq183ee} to obtain the approximate values of $\mathfrak{a}(t)$ and $H(t)$ 
\begin{eqnarray}
	\mathfrak{a}(t) = \mathfrak{a}(T) \left[1-\ds \frac{1}{2} H^2(T) R^2\right]+\mathcal{O} \left[H^3(T)\right]\,,\label{eq187b}\\
	H(t) = H(T) + \mathcal{O} \left[H^3(T)\right]\,.\label{eq187a}
\end{eqnarray}
From here onwards, we shall denote $\mathfrak{a}(T) \equiv \mathfrak{a}$, $H(T) \equiv H$ and  $\kappa \equiv {k}/{\mathfrak{a}^2(T)}$, where $\kappa$ is called the Gaussian curvature of space.

We substitute equations \eref{eq158dd}, \eref{eq177i}, \eref{eq182c}, \eref{eq187b}, and \eref{eq187a} in equation \eref{eq185a} to obtain the time-time component ${G}_{TT}(T,R)\equiv G_{TT}$ as 
\begin{eqnarray}\label{eq187ab1}
\hspace{-2cm} G_{TT}=\left\{\left(R^2 H^2-x^2\right)+\ds \frac{\kappa m^2 \left(x+1\right)^3 \left[\left(x-1\right) x+4 R^2 H^2\right]}{8 \left(x-1\right)^2 x \left(R^2 H^2-2\right)^2}\right\}\left[1 - \ds \frac{\partial \chi\left(T,R\right)}{\partial T}\right]^2\,.
\end{eqnarray}
We substitute for ${\partial \chi(T,R)}/{\partial T}$ from equation \eref{eq187a1} in equation \eref{eq187ab1}, the result obtained is given expansion with respect to the small parameters $m/R$ (post-Newtonian) and $H R/c$. It leads to
\begin{equation}\label{eq188}
	\hspace{-1cm}	G_{TT}=-\left[1-\ds \frac{2 m}{R}- H^2R^2-\dot{H}R^2 \left(1+\frac{4m}{R}\right) +\ds \frac{\kappa }{4}\left(\ds \frac{m}{R}-2 \dot{H}R^2\right)R^2\right]\,,
\end{equation}
where we have dropped all the residual higher order terms of order $m^2/R^2, H^3, \dot{H}^2$, etc.

Similarly, we substitute equations \eref{eq158dd}, \eref{eq182c}, \eref{eq187b}, and \eref{eq187a} in equation \eref{eq185} to obtain the space-space component ${G}_{RR}(T,R)\equiv G_{RR}$ as 
\begin{eqnarray}\label{eq187ab2}
	 G_{RR}= \ds \frac{1}{\left[x-\ds \frac{\kappa m^2 \left(x+1\right)^3 \left(5 x-1\right)}{64 \left(x-1\right)^2 x \left(1-\frac{1}{2} R^2 H^2\right)^2}\right]^2-R^2 H^2}\,.
\end{eqnarray}
\Eref{eq187ab2} must be expanded with respect to the small parameters $m/R$ (post-Newtonian) and $H R/c$. It leads to
\begin{eqnarray}
	\hspace{-1cm}	G_{RR}=1+\ds \frac{2 m}{R} + H^2R^2 \left(1+\frac{4m}{R}\right) +\kappa \left(1 +\ds \frac{m}{4R}+3 H^2 R^2\right)R^2\,,\label{eq189}
\end{eqnarray}
where we have dropped all the residual higher order terms of order $m^2/R^2, H^3, \dot{H}^2$, etc. Equations \eref{eq188} and \eref{eq189} represent the post-Friedmanian expansion of the cosmological McVittie's metric in the local coordinates $(T,R)$.  
\section{Equations of the motion of a test particle}\label{sc4}
The position vector of a test particle is given by vector $\vec{R}\equiv(X^i)$. The equation of motion for the freely falling test particle is the time-like geodesic
\begin{eqnarray}
	\ds \frac{{\rmd}^2X^i}{{\rmd} T^2} + \Gamma^i{}_{\mu\nu} 	\ds \frac{{\rmd} X^\mu}{{\rmd} T}\ds \frac{{\rmd} X^\nu}{{\rmd} T} - \Gamma^0{}_{\mu\nu} \ds \frac{{\rmd} X^\mu}{{\rmd} T}\ds \frac{{\rmd} X^\nu}{{\rmd} T}\ds \frac{{\rmd} X^i}{{\rmd} T} =0 \,,\label{eq192}
\end{eqnarray}
where $X^\mu = (T, X^i), \: V^i = {{\rmd} X^i}/{{\rmd} T}$ is the velocity of the particle with respect to the local coordinates and the coordinate time $T$ has been used as a parameter along the particle world line. The non-vanishing components of Christoffel's symbols are computed by making use of the metric tensor given by equations \eref{eq188} and \eref{eq189}, listed in \tref{table1}. 
\begin{table}[ht]
	\caption{\label{table1}Non-vanishing components of Christoffel's symbols.} 
	\begin{indented}
		\lineup
		\item[]\fl
		{\tiny	\begin{tabular}{@{}*{7}{l}}
				\br  
				Christoffel's symbols & Terms without cosmological effects & Terms with $H$ and $\dot {H}$ & Terms with curvature $k$ \cr
				\mr 
				$\Gamma^0{}_{00} $ & $0$ &  $\sim H^3$ (negligible) & $\sim H^3$ (negligible) \cr
				$\Gamma^i{}_{0j} $ & $0$ &  $0$ &  $-\kappa H \left(1-\ds \frac{2m}{R}
				\right) R^i R^j$ \cr
				$\Gamma^0{}_{jk} $ & $0$ &  $0$ &  $- \kappa H  \left(1+\ds \frac{2m}{R}+ \ds \frac{4m^2}{R^2} \right) R^j R^k $ \cr
				$\Gamma^0{}_{0j} $ & $\left(1+\ds \frac{2m}{R}
				\right)\ds \frac{m}{R^2} N^j$ & $-\left[\left(1+\ds \frac{m}{R}\right)H^2
				+ \left(1+\ds \frac{3m}{R}\right) \dot{H}\right] R^j$ & $ \ds \frac{\kappa}{8}m N^j$ \cr
				$\Gamma^i{}_{00} $ & $\left(1-\ds \frac{2m}{R}
				\right)\ds \frac{m}{R^2} N^i$ & $ -\left[\left(1-\ds \frac{m}{R}\right)H^2+ \dot{H}\right]R^i$ & $ -\ds \frac{7\kappa}{8} m N^i$ \cr
				$\Gamma^i{}_{jk} $ & $\ds \frac{m}{R^2}\left[2 \delta _{jk}-\left(3+\ds \frac{2m}{R}\right) N^j N^k\right]N^i$ & $H^2 \left(\delta _{jk}+\ds \frac{ m}{R} N^j N^k \right)R^i$ & $\kappa \left[\left(1-\ds \frac{15m}{4R}\right) \delta _{jk} +\ds \frac{23 m}{8R} N^j N^k \right]R^i $ \cr
				\br
		\end{tabular}}
	\end{indented}
\end{table}
We use the Christoffel's symbols from \tref{table1} to compute the equation of motion of the test particle explicitly. The equation has the form of Newton's second law and reads
\begin{eqnarray}
	\ddot{\vec{R}}= \vec{F}_{N} +\vec{F}_{pN} + \vec{F}_{H} + \vec{F}_{K}\, .\label{eq193d}
\end{eqnarray} 
Here, in equation~\eref{eq193d},
\begin{eqnarray}
	\hspace{-1cm}	\vec{F}_{N}= - \ds \frac{G M }{R^2}~\vec{N}\, ,\label{eq193e}\\
	\hspace{-1cm}	\vec{F}_{pN} =  \ds \frac{m}{R^2}\left[ \left(\ds \frac{2 G M}{R}  + 3 V_R^2 -2 V^2 \right)\vec{N} + 2 V_R\vec{V}\right]\,, \label{eq193f}\\
	\hspace{-1cm} \vec{F}_{H} = H^2 \left[ \vec{R} -\ds \frac{1}{c^2} \left(G M \vec{N} + 2 R V_R \vec{V}+ V^2 \vec{R}\right)\right] + \dot{H}\left[ \vec{R} - \ds \frac{2}{c^2} \left(R V_R \vec{V}\right)\right]\,, \label{eq193g}\\
	\hspace{-1cm}	\vec{F}_{K}=\kappa \left(\ds\frac{7 G M}{8R}-V^2\right)\vec{R}\,,\label{eq193h}
\end{eqnarray}
where $ \vec{N} = {\vec{R}}/{R}$ is the unit vector, $\vec{V} = \dot{\vec{R}}$ is the velocity of the particle, $V_R=\vec{N}\cdot\vec{V}$ is the radial component of the velocity, $\vec{F}_{N}$ is the Newtonian force, $\vec{F}_{pN}$ is the post-Newtonian perturbing force, $\vec{F}_{H}$ is the perturbing force due to the Hubble expansion and $\vec{F}_{K}$ is the perturbing force due to the Gaussian curvature of the cosmological space.
\subsection{Unperturbed orbits}\label{sbscPr} 
In absence of any perturbing force the equation of motion, equation~\eref{eq193d}, reduces to the Newtonian form with the Newtonian force that is given by equation~\eref{eq193e}.
We consider the elliptic motion of the particle whose Keplerian orbit is given by the radius vector~$R$. The reason we consider the elliptical orbits of test particles is that we are interested in the analysis of stability of the orbital motion of particles in gravitationally bound systems subject to cosmological expansion. The elliptic orbit is defined by equation
\begin{eqnarray}\label{eq195a}
	R= \ds \frac{p}{1+e \cos f} \,,
\end{eqnarray}
where $p=a(1-e^2)$ is the focal parameter of the orbit, $a$ is the semi-major axis, $e$ is the eccentricity and $f$ is a true anomaly (the angle between the directions to the particle and to the pericenter of its orbit) which is a function of time $T$. The relation of $f$ with time $T$ are given by transcendental equations  
\begin{eqnarray}\label{eq198a}
	\tan \ds \frac{f}{2} = \left(\ds \frac{1+e}{1-e}\right)^{\frac{1}{2}} \tan \ds \frac{E}{2}\,,
\end{eqnarray} 
\begin{eqnarray}\label{eq197a}
	E - e \sin E = l \,,
\end{eqnarray}
where $l$ is mean anomaly which is a linear function of time $T$
\begin{eqnarray}
	l = \mathfrak{M} + n \left(T - T_0\right)\, ,\label{eq200}
\end{eqnarray}
and $\mathfrak{M}$ is the mean anomaly at the epoch. The mean motion $n$ in equation \eref{eq200} is related to the semi-major axis $a$ by Kepler's third law $n^2 a^3 = G M$. 

In what follows it is convenient to introduce an orbital frame characterized by three unit vectors $\vec{e}_R$, $\vec{e}_T$ and $\vec{e}_W$, 
\begin{eqnarray}
	 \vec{e}_R = \vec{N}\,, \qquad  \qquad
	\vec{e}_T = \vec{e}_R \times \vec{e}_W \,, \qquad \qquad
	\vec{e}_W = \ds \frac{\vec{R} \times \vec{V}}{\left|\vec{R} \times \vec{V}\right|}\,, \label{eq202h}
\end{eqnarray}
where $\vec{e}_R$, $\vec{e}_T$ lie in the plane of the orbital motion and $\vec{e}_W$ is normal to it. The velocity $\vec{V}$ of the test particle is decomposed with respect to these unit vectors as follows:
\begin{eqnarray}\label{eq204a1}
	 \vec{V}   = V_{R}\;\vec{e}_R + V_{T}\;\vec{e}_T\,.
\end{eqnarray}
The components $V_R$ and $V_T$ in equation \eref{eq204a1} are expressed in terms of the orbital elements as \cite{brumbergbook1991}
\numparts
\begin{eqnarray}
	V_{R} =\left[GM \left( \ds \frac{2}{R}- \ds \frac{p}{R^2}-\ds \frac{1}{a}\right)\right]^{\frac{1}{2}} = \sqrt{\ds \frac{GM}{p}}e\sin f\,,\label{eq207b}\\
	V_{T} = \ds \frac{\sqrt{GMp}}{R}=\sqrt{\ds \frac{GM}{p}}\left(1+e\cos f\right)\,.\label{eq206b}
\end{eqnarray}
\endnumparts
The magnitude $V$ of the velocity $\vec{V}$ is given by
\begin{eqnarray}
	V = \left[GM \left(\ds \frac{2}{R}-\ds \frac{1}{a}\right)\right]^{\frac{1}{2}}\,.\label{eq205}
\end{eqnarray}
In the presence of the perturbing forces the orbital elements of the elliptic orbit become functions of time, which are known under the name of osculating elements \cite{brumbergbook1991}. The time dependence of the osculating elements can be found by solving the system of the first order ordinary differential equations which are discussed  below.
\subsection{Osculating elements}\label{sbscOs}
\begin{figure}
	\centering
	\includegraphics[scale=0.17]{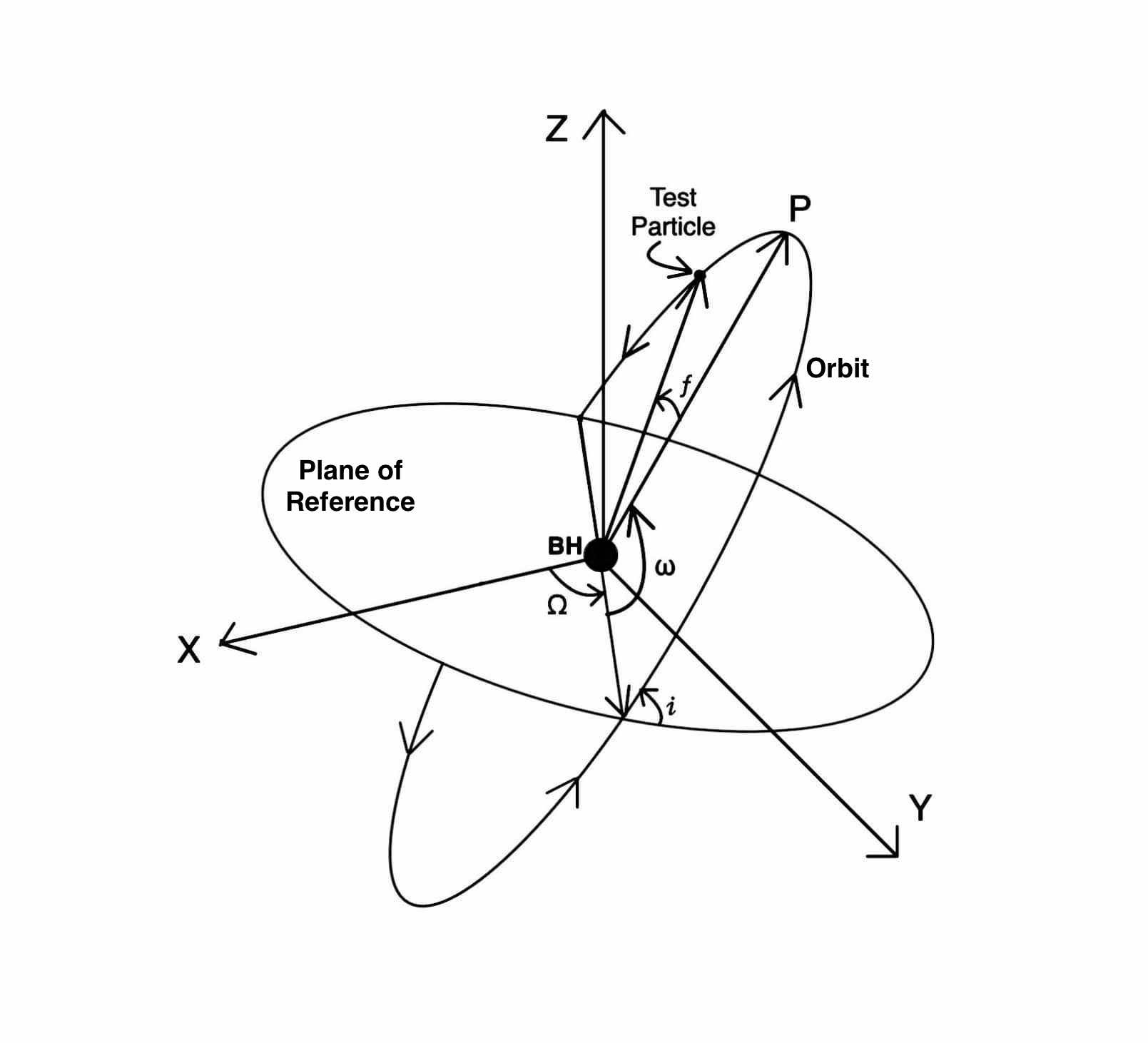}
	\caption{Orbit of a test particle around a black hole (BH). The notations used in the figure are explained in subsection \ref{sbscOs}.\label{Orbit}}  
\end{figure}
We study the effects of the perturbing forces on the six orbital elements of an elliptical orbit shown in \fref{Orbit}, that are:

\noindent -- $a$ is the semi-major axis defined as half the distance between the apoapsis and periapsis. It measures the size of the ellipse.

\noindent -- $e$ is the eccentricity that describes the shape of the ellipse, that is, how much the ellipse is elongated compared to a circle.

\noindent -- $i$ is the orbital inclination (the angle between the $xy$ plane and the orbital plane of the particle) that defines the orientation of the particle's orbit in space.

\noindent -- $\Omega$ is the longitude of the ascending node (the angle in the $xy$ plane between the the $x$-axis and the line of nodes).

\noindent -- $\omega$ is the argument of pericenter defined as the angle between the pericenter (P) and the ascending node of the orbit.

\noindent -- $\mathfrak{M}$ is the mean anomaly at the epoch.

Due to the presence of the perturbing force the orbital elements experience changes which can be found by integrating the system of equations \cite{brumbergbook1991}
\numparts
\begin{eqnarray}
		 \ds \frac{{\rmd} a}{{\rmd} T} = \ds \frac{2}{n \left(1-e^2\right)^{\frac{1}{2}}} \left(S e \sin f + T \ds \frac{p}{R}\right)\,,\label{eq202a}\\
		 \ds \frac{{\rmd} e}{{\rmd} T} = \ds \frac{\left(1-e^2\right)^{\frac{1}{2}}}{n a} \left[S \sin f + T\left(\cos f+\cos E\right)\right]\,,\label{eq202b}\\
		\ds \frac{{\rmd} \omega}{{\rmd} T} = -\cos i \ds \frac{{\rmd} \Omega }{{\rmd} T}+\ds \frac{\left(1-e^2\right)^{\frac{1}{2}}}{nae}\left[-S \cos f + T \left(1+ \ds \frac{R}{p} \right)\sin f \right]\,,\label{eq202e}\\
		 \ds \frac{{\rmd} \mathfrak{M}}{{\rmd} T} = -\left(1-e^2\right)^{\frac{1}{2}}\left(\ds \frac{{\rmd} \omega }{{\rmd} T}+\cos i \ds \frac{{\rmd} \Omega}{{\rmd} T} \right) - S \ds \frac{2R}{na^2} \,,\label{eq202f}\\
    	\ds \frac{{\rmd} i}{{\rmd} T} = \ds \frac{R \cos (f+\omega)}{n a^2 \left(1-e^2\right)^{\frac{1}{2}}} W\,,\label{eq202c}\\
		 \ds \frac{{\rmd} \Omega}{{\rmd} T} = \ds \frac{R \sin (f+\omega)}{n a^2 \left(1-e^2\right)^{\frac{1}{2}}\sin i} W\,,\label{eq202d}
\end{eqnarray}
\endnumparts
where $S$, $T$, $W$ are the components of the perturbing force along the radial, transverse and normal directions to the plane of motion
\begin{eqnarray}
S = \vec{e}_R\cdot\vec{F},\qquad T = \vec{e}_T\cdot\vec{F}, \qquad W = \vec{e}_W\cdot \vec{F}\,.\label{eq213aaa1a1}
\end{eqnarray} 
Specific perturbations in each of the osculating elements caused by a certain type of the perturbing forces $(\vec{F}_{pN}, \vec{F}_{H}, \vec{F}_{K})$ are denoted by the label $pN$, $H$, $K$ corresponding to the force.

The $W$ components $W_{pN}$, $W_H$, $W_K$ (which are normal to the plane of motion) of the perturbing forces $\vec{F}_{pN}$, $\vec{F}_{H}$, and $\vec{F}_{K}$ respectively are zero. 
\begin{eqnarray}
	\vec{e}_W\cdot\vec{F}_{pN} =
	\vec{e}_W\cdot\vec{F}_H =
	\vec{e}_W\cdot \vec{F}_K = 0 \,.\label{eq248a1}
\end{eqnarray}
As a consequence, it immediately follows from equation \eref{eq202c} and equation \eref{eq202d} that the inclination $i$ and the ascending angle $\Omega$ are constant.
It means that the orbital plane does not change its spatial orientation. In what follows we analyze the secular behavior of the remaining four osculating elements. 

The osculating elements have secular and periodic perturbations under the influence of the perturbing force. Though the periodic perturbations are interesting from a theoretical point of view, they are too small in most of the practical situations. Therefore, in what follows, we primarily focus on calculations of the secular variations of the osculating elements. This is achieved by applying the time averaging technique \cite{brumbergbook1991}. 
\subsection{The averaging technique}\label{sbscAv}
There are two time scales -- short and long -- in the problem under consideration. The short time scale corresponds to the orbital motion of the test particle around central black hole. It is characterized by the orbital period $P_b$. The long time scale corresponds to the expansion of the universe and is characterized by the Hubble time $T_H=1/H$.  We are assuming that $P_b \ll T_H$. The time argument associated with orbital motion of the test particle is called a fast independent variable as it changes on the short time scale of the orbital period $P_b$. The true anomaly $f$ as well as the eccentric anomaly $E$, and the mean anomaly $l$, are examples of the fast variables as they change significantly over one orbital period of motion of the test particle. The time argument associated with the Hubble  expansion is called  a slow independent variable as it changes on the Hubble time $T_H$. The examples of the slow variables are scale factor $\mathfrak{a}$, the Hubble parameter $H$ and its time derivative $\dot{H}$ because they change noticeably only over the period of the Hubble time. Therefore, we consider the slow variables $\mathfrak{a}$, $H$ and $\dot{H}$ as constant over one orbital period.

The perturbing force is a function of both types of the variables, let say, $F(T)\equiv F\left[\alpha(T),\beta(T)\right]$, where $\alpha(T)$ is a set of slow variables and $\beta(T)$ is a set of fast variables. The average value of the function $F(T)$ is defined \cite{brumbergbook1991} as a time integral performed over one orbital period with respect to a fast variable of the true anomaly $f$,
\begin{eqnarray}\label{eq239c1}
	\left\langle F(T) \right\rangle = \ds \frac{1}{P_b} \int_{-P_b/2}^{P_b/2} F(T)\,  {\rmd} T = \ds \frac{1}{2 \pi}\int_{-\pi}^{\pi} F\left[\alpha(T),\hat{\beta}(f)\right] \,\ds\frac{{\rmd}T}{{\rmd} f}\, {\rmd} f + ... \,,
\end{eqnarray}
where $\hat{\beta}(f)\equiv \beta \left[T(f)\right]$. Equation \eref{eq239c1} will be used in calculation of the average value of the perturbing force. In a particular case, when the function $F(T)$ $=$ $\left({R}/{a}\right)^q {\rm exp}({\rm i}sf); ~q, s \in \mathbb{N}$ (integers), the average value of this function is given by \cite{brumbergbook1991}
\begin{eqnarray}
	\left\langle \left(\ds \frac{R}{a}\right)^q \cos sf \right\rangle= X_0^{q,s}(e)\,\qquad,  \qquad \left\langle\left(\ds \frac{R}{a}\right)^q \sin sf\right\rangle =0\,,\label{eq204a}
\end{eqnarray}
where $q$, $s$ are integers. $X_0^{q,s}(e)$ are the Hansen coefficients depending only on the eccentricity $e$ of the orbit of the test particle and defined by \cite{brumbergbook1991}
 \begin{eqnarray}\label{eq204b}
\hspace{-2.3cm}	X_0^{q,s}(e) =  & \left(\ds \frac{e}{2}\right)^{|s|} \ds \frac{\left(-q-1-|s|\right)_{|s|}}{(1)_{|s|}}\,\nonumber\\
	& 
\hspace{-2cm}	\times \cases{
		{}_2F_1 \left(\ds \frac{|s|-q-1}{2},\ds \frac{|s|-q}{2};1+|s|;e^2\right) \quad & $q\geq|s|-1$\\
		\left(1+\beta^2\right)^{|s|-q-1}{}_2F_1\left(|s|-q-1,-q-1;1+|s|;\beta^2\right) \quad & $|s|-1>q\geq-1$\\
		~~~~~~~~~~~~~~~~~~~~~~~~~~~~ 0 \quad & $-1>q\geq-|s|-1$\\
		\left(1-e^2\right)^{q+3/2}{}_2F_1 \left(\ds \frac{q+|s|+2}{2},\ds \frac{q+|s|+3}{2};1+|s|;e^2\right) \quad & $-|s| -1 > q $\,,}\nonumber\\
\end{eqnarray}
with $\beta = \ds \frac{e}{1+\left(1-e^2\right)^{1/2}}$. Some particular values of the Hansen coefficients used in our calculations are given in \ref{scap}. 
\section{Post-Newtonian perturbations of osculating elements due to the central black hole}\label{scpN}
Post-Newtonian perturbations of the osculating elements in a two-body problem were studied by Brumberg \cite{brumbergbook1991}. 
The radial $S_{pN}$ and transverse $T_{pN}$ components of the perturbing force are
\begin{eqnarray}
	 S_{pN} = \vec{e}_R \cdot \vec{F}_{pN}=m n^2\ds \frac{a^2}{R^2} \left(-3+8\ds \frac{a}{R}-5\ds \frac{a p}{R^2}\right)\,,\label{eq208a}
\end{eqnarray}
\begin{eqnarray}
	 T_{pN} =\vec{e}_T\cdot\vec{F}_{pN}= 2m n^2 e \ds \frac{ a^3}{R^3} \sin f \,.\label{eq210a}
\end{eqnarray}
The perturbation of osculating elements $a$, $e$, and $\omega$ due to the post-Newtonian approximation are determined by substituting the components of the perturbing force $S_{pN}$ from equation \eref{eq208a} and $T_{pN}$ from equation \eref{eq210a} into equations \eref{eq202a}-\eref{eq202e}.
The differential equation for the the post-Newtonian perturbation of the semi-major axis $a$ is
\begin{eqnarray}\label{eq214a}
	 \ds \frac{{\rmd} a}{{\rmd} T} = \ds \frac{2mne a p}{\left(1-e^2\right)^{\frac{3}{2}}R^2}\left(-3+8\ds \frac{a}{R}-3\ds \frac{a p}{R^2}\right)\sin f\,.
\end{eqnarray}
Applying equation \eref{eq204a} and the Hansen coefficients from \ref{scap} to calculate the average value, we get 
\begin{eqnarray}
	\ds \left\langle \frac{{\rmd a_{pN}}}{{\rmd} T}  \right\rangle= 0 \,,\label{eq216a}
\end{eqnarray}
which means that in the local coordinates the average value of the post-Newtonian perturbation of the semi-major axis vanishes, and the mean value of the semi-major axis is constant.

The differential equation for the post-Newtonian perturbation of eccentricity $e$ is
\begin{eqnarray}\label{eq217a}
	\ds \frac{{\rmd} e}{{\rmd} T} =\ds \frac{m n p}{\left(1-e^2\right)^{\frac{1}{2}}R^2} \left(-5+8 \ds \frac{a}{R}-3\ds \frac{a p }{R^2} \right)\sin f\,.
\end{eqnarray}
Calculating the average value of equation \eref{eq217a} by applying equation \eref{eq204a} and the results of \ref{scap}, we get 
\begin{eqnarray}
	\left\langle	\ds \frac{{\rmd} e_{pN} }{{\rmd} T}\right\rangle = 0\,,\label{eq218a}
\end{eqnarray}
which means that in the local coordinates the average value of the eccentricity of the particle's orbit is not perturbed by the post-Newtonian force and remains constant.

The differential equation for the post-Newtonian perturbation of the argument of the orbital pericenter ${\omega}$ is 
\begin{eqnarray}
	\ds \frac{{\rmd}\omega}{{\rmd} T} =  \ds \frac{m n p}{e^2\left(1-e^2\right)^{\frac{1}{2}}  R^2 } \left[-5+12\ds  \frac{a}{R}+\ds \frac{p}{R}\left(1-11 \ds \frac{a}{R}+3\ds \frac{a p}{R^2}\right)\right].\label{eq219b1}
\end{eqnarray}
The average value of the time rate of change of the argument of pericenter ${\omega}$, that is equation \eref{eq219b1}, is obtained with the help of equations \eref{eq204v}-\eref{eq204s}. It yields 
\begin{eqnarray}
	\left\langle \ds \frac{{\rmd} \omega_{pN} }{{\rmd} T} \right\rangle =\ds \frac{3 m n}{a\left(1- e^2\right)} \,,\label{eq219b2}
\end{eqnarray}
which is, of course, a well-known post-Newtonian precession of the elliptic orbit in two-body problem \cite{1960_S}.

The differential equation for the post-Newtonian perturbation of the mean anomaly at epoch $\mathfrak{M}$ is 
\begin{eqnarray}
		\ds \frac{{\rmd} \mathfrak{M}}{{\rmd}T} = -\left(1-e^2\right)^{\frac{1}{2}}\ds \frac{{\rmd} \omega }{{\rmd} T}
	+ \ds \frac{2 m n}{R} \left(3-8 \ds \frac{a}{R}+5\ds \frac{a p}{R^2}\right)
	.\label{eq219b1a}
\end{eqnarray}
We figure out the average value of the rate of change of the mean anomaly at epoch $\mathfrak{M}$, that is equation \eref{eq219b1a}, with the help of equations \eref{eq204c}-\eref{eq204s}. The result is
\begin{eqnarray}
	\left\langle\ds \frac{{\rmd}\mathfrak{M}_{pN}}{{\rmd}T}  \right\rangle=  \ds \frac{3 m n}{a } \left[2 - \ds \frac{ 3 }{\left(1-e^2\right)^{\frac{1}{2}}}\right]
	\,.\label{eq219b1b}
\end{eqnarray}
Equation \eref{eq219b1b} explains the change in the frequency of the mean orbital motion due to the post-Newtonian approximation.
\section{Perturbations of osculating elements due to the Hubble expansion}\label{scH}
Perturbations of osculating elements of the test particle's orbit due to the Hubble expansion are computed by making use of the perturbing force $\vec{F}_H$ given by equation~\eref{eq193g}. We calculate the radial force $S_H$, and transverse force $T_H$, components of the perturbing force, and get 
\begin{eqnarray}
\hspace{-1.5cm}	S_{H} = H^2 R\left[1 -\ds \frac{m}{R} \left(7 -3\ds \frac{R}{a}-2 \ds \frac{p}{R}\right)\right]
	+  \dot{H} R \left[1  - \ds \frac{2m}{R} \left(2 - \ds \frac{R}{a} - \ds \frac{p}{R}\right)\right]\,,\label{eq226a1a}
\end{eqnarray}
\begin{eqnarray}
\hspace{-1.5cm}	T_H  = -2 m  \left(H^2 + \dot{H}\right) e \sin f\,. \label{eq228a2a}
\end{eqnarray}
The expression for $\dot{H}$ in terms of Hubble parameter $H$ is given by
\begin{eqnarray}
	\dot{H} = -H^2\left(1+\mathfrak{q}\right) \,, \label{eq228d1}
\end{eqnarray}
where $\mathfrak{q} = - {\ddot{\mathfrak{a}}\mathfrak{a}}/{\dot{\mathfrak{a}}^2}$ is the dimensionless deceleration parameter \cite{weinberg_2008}. It describes the rate of change in Hubble's expansion. A positive $\mathfrak{q}$ indicates $\ddot{\mathfrak{a}}<0$, which means the universe’s expansion is slowing down, or decelerating. A negative $\mathfrak{q}$ signifies $\ddot{\mathfrak{a}}>0$, which means the universe’s expansion is accelerating. This is typically attributed to the influence of dark energy, which drives the accelerated expansion. Current observations indicate a negative value of $\mathfrak{q}$ \cite{2011ENews..42f..20V}. When $\mathfrak{q}$ is zero, $\dot{\mathfrak{a}}={\rm constant}$, which means the universe’s expansion rate is constant. This implies a balance between the forces causing expansion and those causing deceleration, resulting in a steady rate of expansion. This is typically for the de Sitter universe.

We can see from equation \eref{eq228d1} that $\dot{H}$ is of the same order as $H^2$, that is, $\dot{H} \sim H^2$. Therefore, it is instrumental to re-write equations \eref{eq226a1a}, \eref{eq228a2a} in terms of deceleration parameter $\mathfrak{q}$,
\begin{eqnarray}
	\hspace{-1cm}	S_{H} =- \mathfrak{q} H^2 R + m H^2\left[ 2 \mathfrak{q} \left(2 - \ds \frac{R}{a} - \ds \frac{p}{R}\right)-\left(3 -\ds \frac{R}{a}+4 \ds \frac{p}{R}\right)\right]\,,\label{eq226}
\end{eqnarray}
\begin{eqnarray}
	\hspace{-1cm}	T_H  = 2 m \mathfrak{q} e H^2 \sin f\,. \label{eq228}
\end{eqnarray}

The Hubble parameter $H$, $\dot{H}$, the deceleration parameter $\mathfrak{q}$ in equations \eref{eq226}, \eref{eq228} and the scale factor $\ds {\mathfrak{a}}$ are series expanded about the present epoch $T=T_0$ as
\numparts
\begin{eqnarray}
	\hspace{-1.5cm}	H = H_0 + \dot{H}_0 (T-T_0) + \mathcal{O}  \left[ \left(T-T_0\right)^2\right]\,, \label{eq228a}\\
	\hspace{-1.5cm}	\dot{H} = \dot{H}_0 + \mathcal{O}  \left[ \left(T-T_0\right)\right]\,, \label{eq228b}\\
	\hspace{-1.5cm}	\ds {\mathfrak{q}} = {\mathfrak{q}_0}+\dot{\mathfrak{q}}_0 \left(T-T_0\right)+\mathcal{O} \left[ \left(T-T_0\right)^2\right]\,, \label{eq228d}\\
	\hspace{-1.5cm}	\ds {\mathfrak{a}} = {\mathfrak{a}_0}+\dot{\mathfrak{a}}_0 \left(T-T_0\right)+\mathcal{O} \left[\left(T-T_0\right)^2\right]\,, \label{eq228c}
\end{eqnarray}
\endnumparts
where $H_0$, $\dot{H}_0$, $\mathfrak{q}_0$ and ${\mathfrak{a}_0}$ are the constant values of the parameters taken at $T=T_0$.  
The perturbation of osculating elements $a$, $e$, $\omega$ and $\mathfrak{M}$ due to Hubble's expansion are determined below.
\subsection{Semi-major axis $a$.}\label{sma}
The differential equation for the time evolution of the semi-major axis $a$ after substituting $S_H$ from equation \eref{eq226} and $T_H$ from equation \eref{eq228} into equation \eref{eq202a} reads 
\begin{eqnarray}
	\hspace{-1cm}\frac{{\rmd a_{H}}}{{\rmd} T}  = \ds \frac{2e}{n\left(1-e^2\right)^{\frac{1}{2}}}H^2 \left\{-\mathfrak{q} R + m \left[2\mathfrak{q} \left(2-\ds \frac{R}{a} \right)
	- \left(3-\ds \frac{R}{a}\right)\right]\right\}\sin f\,.\label{eq231b}
\end{eqnarray}
We calculate the average of each term of equation \eref{eq231b} over a time period $P_b$ by making use of equation \eref{eq204a}, the Hansen coefficients from \ref{scap}, and the first terms from equations \eref{eq228a}-\eref{eq228c}. It leads to
\begin{eqnarray}
	\ds \left\langle \frac{{\rmd a_{H}}}{{\rmd} T}  \right\rangle= 0  \,.\label{eq233b}
\end{eqnarray}
\Eref{eq233b} concludes that the average value of the semi-major axis is not affected by the perturbing force caused by the Hubble expansion to the order of $H_0^2$ and is constant, that is, the average value of semi-major axis of the particle's orbit is not subject to cosmological expansion to the order of $H_0^2$. This conclusion is fully consistent with Einstein's principle of equivalence and agrees with the results obtained by the other researchers studying the influence of the cosmological expansion on the elliptical orbits of planets and binary stars \cite{mcvittie1933, carrera2010, bolen2001, Kopeikin_2012PhRvD}.

However, we can proceed similar to the above method to calculate the perturbation of semi-major axis of the particle's orbit by taking into account the second and further terms in the expansion of equations \eref{eq228a}-\eref{eq228c}. We can see in that case the average values for the terms of equation \eref{eq231b} will contain a factor like $\left\langle H_0^{2+n} T^n \sin f  \right\rangle \neq 0$, for any value of $n\geq1$. Here time $T$ should be expressed in terms of true anomaly $f$ through Kepler's equation \eref{eq197a}. This indicates that, in general, we might expect $\left\langle {\rmd  a_{H}}/{{\rmd} T}  \right\rangle \neq 0 $ when more terms in the expansions shown in equations \eref{eq228a}-\eref{eq228c} are used to calculate the average values. However, as demonstrated in Appendix C, this expectation is not supported by the analysis of the asymptotic behavior of $\left\langle {\rmd  a_{H}}/{{\rmd} T}  \right\rangle $ for very large values of time $T$.
\subsection{Eccentricity $e$.}\label{ecc}
The differential equation for the eccentricity $e$ due to the Hubble expansion is obtained after substituting the radial force $S_H$ from equation \eref{eq226} and the transverse force $T_H$ from equation \eref{eq228} to equation \eref{eq202b}. It yields
\begin{eqnarray}
	\hspace{-1cm}	\frac{{\rmd} e_{H} }{{\rmd} T} =  \ds \frac{\left(1-e^2\right)^{\frac{1}{2}}}{na}H^2 \left\{-\mathfrak{q} R + m \left[4 \mathfrak{q} \left(1- \ds \frac{R}{a} \right) - \left( 3-\ds \frac{R}{a}\right)\right]\right\}\sin f\,.\label{eq231bb}
\end{eqnarray}
We determine the average value of each term in equation \eref{eq231bb} over a time period $P_b$ by making use of equation \eref{eq204a}, the results from \ref{scap}, and first terms from equations \eref{eq228a}-\eref{eq228c}. It leads to
\begin{eqnarray}
	\left\langle	\ds \frac{{\rmd} e_{H} }{{\rmd} T}\right\rangle = 0\,.\label{eq237b}
\end{eqnarray}
\Eref{eq237b} concludes that $ \left\langle e_{H} \right\rangle$ is constant, that is, the average value of the eccentricity of the particle's orbit is not affected by cosmological expansion to the order of $H_0^2$. This conclusion also agrees with the results obtained by the other researchers \cite{mcvittie1933, bolen2001}. 

However, we can proceed similar to the above method for semi-major axis to calculate the perturbation of eccentricity of the particle's orbit by taking into account the second and further terms in the expansion shown in equations \eref{eq228a}-\eref{eq228c}. We can see in that case the average values for the terms of equation \eref{eq231bb} contain terms like $\left\langle  H_0^{2+n} T^n \sin f  \right\rangle \neq 0$, for any value of $n\geq1$. Here time $T$ should be expressed in terms of true anomaly $f$ through Kepler's equation \eref{eq197a}. The averaging of these terms might lead to  $\left\langle {\rmd  e_{H}}/{{\rmd} T}  \right\rangle \neq 0$. However, this behavior of  $\left\langle {\rmd  e_{H}}/{{\rmd} T}  \right\rangle$ is not supported by the asymptotic analysis for very large values of time $T$, as demonstrated in Appendix C.
\subsection{Argument of pericenter $\omega$.}
The differential equation for the argument of pericenter  $\omega$ after substituting the radial force $S_H$ from equation \eref{eq226} and the transverse force $T_H$ from equation \eref{eq228} to equation~\eref{eq202e} becomes
 \begin{eqnarray}\label{eq238b}
	\hspace{-1.8cm}	\frac{{\rmd} \omega_{H} }{{\rmd} T} =  
	\ds \frac{\left(1-e^2\right)^{\frac{1}{2}}}{nae} & H^2 \left\{ \left[\mathfrak{q} R   -  m \left[2 \mathfrak{q} \left[2 -\left(1 - e^{2}\right)  \ds \frac{a}{R} -\ds \frac{R}{a} \right]  - \left(3  - \ds \frac{R}{a} \right)\right]\right] \cos f 
	\right. \nonumber\\
&	\left.
\hspace{2cm}	+ \ds \frac{m \mathfrak{q} e }{\left(1-e^{2}\right)} \left[ \left(1-e^{2}\right)+ \ds \frac{R}{a} \right]\left(1- \cos 2f\right) \right\}\,.
\end{eqnarray}
After computing the average value of each term in equation \eref{eq238b} over a time period of one orbital revolution $P_b$ by making use of equations \eref{eq204k}-\eref{eq204c}, we get the time rate of change for the average value of the argument of pericenter $\left\langle \dot{\omega}_{H} \right\rangle$ as
\begin{eqnarray}	
	\left\langle \ds \frac{{\rmd} \omega_{H} }{{\rmd} T} \right\rangle= -\ds \frac{3 \left(1-e^2\right)^{\frac{1}{2}}}{2n}H^2\left[\mathfrak{q} \left(1-\ds \frac{4m}{3a}\right)+\ds \frac{m}{a}\right]\,.\label{eq239ub1}	
\end{eqnarray}
Equation \eref{eq239ub1} describes the influence of the cosmological expansion on the evolution of the orientation of elliptical orbits around a central mass $m$ for any type of the FLRW universe having an arbitrary value of $\mathfrak{q}$. This equation generalizes the results of previous researchers \cite{Mashhoon_2003, Arakida_2013} by accounting for an arbitrary value of the deceleration parameter $\mathfrak{q}$ and the influence of the central mass $m$ on the rate of the cosmological expansion. The equation \eref{eq239ub1} agrees with  the other researchers \cite{Mashhoon_2003, Arakida_2013} for value of $\mathfrak{q}=-1$  in the limit $\left(m/a\right) \rightarrow 0$. Further discussion is done in \sref{scCon}.
\subsection{Mean anomaly at the epoch  $\mathfrak{M}$.}
The differential equation \eref{eq202f} for the mean anomaly at the epoch $\mathfrak{M}$ after substituting for the radial force $S_H$ from equation \eref{eq226} and the transverse force $T_H$ from equation~\eref{eq228} becomes
\begin{eqnarray}
	\hspace{-1.7cm}	\ds \frac{{\rmd} \mathfrak{M}_{H}}{{\rmd} T} = -\left(1-e^2\right)^{\frac{1}{2}} \ds \frac{{\rmd} \omega_{H} }{{\rmd} T}+ \ds \frac{2}{n}H^2&\left\{ \mathfrak{q}\left[\ds \frac{R^2}{a^2} + \ds \frac{2m}{a}\left[ \left(1 - e^{2}\right)-  \ds \frac{R}{a} +  \ds \frac{R^2}{a^2} \right]\right] \right. \nonumber\\
	& \left. 
	\hspace{1.5cm} -\ds \frac{mR}{a^2}\left(1+\ds \frac{R}{a}\right) \right\}.\label{eq240b}
\end{eqnarray}
We calculate the average value of each term in equation \eref{eq240b} over a time period $P_b$ by making use of equations \eref{eq239ub1}, \eref{eq204k}-\eref{eq204d}. It yields the time rate of change for the average value of the mean anomaly at the epoch as
\begin{eqnarray}
	\hspace{-1.6cm}	\left\langle\ds \frac{{\rmd} \mathfrak{M}_{H}}{{\rmd} T}\right\rangle= \ds \frac{3 \left(1-e^2\right)}{2n}&H^2\left\{\mathfrak{q} \left[1-\ds \frac{4m}{3a} + \ds \frac{4 }{3\left(1-e^2\right)}\left(1+\ds \frac{3}{2}e^2+\ds \frac{2m}{a}\right)\right]
	\right. \nonumber\\
	& \left.
	\hspace{1.7cm} + \ds \frac{m}{a} \left(1-\ds \frac{8}{3}\ds \frac{1+e^2}{1-e^2}\right)\right\}\,.\label{eq242lb}
\end{eqnarray}
The perturbation in the mean anomaly at the epoch $\mathfrak{M}$ with time due to the cosmological expansion can be understood as the change in frequency of the mean orbital motion $n$ as we discuss in \sref{scCon}. 
\section{Perturbations of the osculating elements due to the spatial curvature }\label{sck}
We calculate the $S$, $T$, and $W$ components of the perturbing force due to the curvature of space $k$ given by $\vec{F}_K$ in equation \eref{eq193h}. We begin with the $S$ component
\begin{eqnarray}
	S_K = -\kappa n^2 a^3 \left(\ds \frac{9}{8}-\ds \frac{R}{a}\right)\,.\label{eq246}
\end{eqnarray}
where $\kappa \equiv {k}/{\mathfrak{a}^2}$. The transverse component $T_K$ of the perturbing force due to the curvature of space $k$ given by $\vec{F}_K$ in equation \eref{eq193h} is
\begin{eqnarray}
	T_K  = 0 \,.  \label{eq247}
\end{eqnarray} 

The differential equation \eref{eq202a} for the time evolution of the semi major axis $a$ after substituting for $S_K$ and $T_K$ from equations \eref{eq246}, \eref{eq247} yields
\begin{eqnarray}
	\ds  \frac{{\rmd} a_{K} }{{\rmd} T} = -\ds \frac{2 \kappa n a^3 e}{ \left(1-e^2\right)^{\frac{1}{2}}} \left(\ds \frac{9}{8}-\ds \frac{R}{a}\right) \sin f \,.\label{eq250}
\end{eqnarray}
The average of each term in equation \eref{eq250} over a time period $P_b$ by making use of equation \eref{eq204a} and taking $ {1}/{\mathfrak{a}^2} \approx  {1}/{\mathfrak{a}_0^2}$ from equation \eref{eq228c} leads to
\begin{eqnarray}
	\left\langle	\ds  \frac{{\rmd} a_{K} }{{\rmd} T}\right\rangle = 0\,.\label{eq250a}
\end{eqnarray}
\Eref{eq250a} concludes that $\left\langle a_{K} \right\rangle ={\rm const}$.

The differential equation \eref{eq202b} for the time evolution of the eccentricity $e$ after substituting for $S_K$ and $T_K$ from equations \eref{eq246}, \eref{eq247} is
\begin{eqnarray}
\ds \frac{{\rmd} e_{K} }{{\rmd} T} =-  {\kappa na^2\left(1-e^2\right)^{\frac{1}{2}}} \left(\ds \frac{9}{8}-\ds \frac{R}{a}\right)\sin f\,.\label{eq251}
\end{eqnarray}
Taking the average value of each term of equation \eref{eq251} over a time period $P_b$ by making use of equation \eref{eq204a} and taking $ {1}/{\mathfrak{a}^2} \approx  {1}/{\mathfrak{a}_0^2}$ from equation \eref{eq228c} yields
\begin{eqnarray}
	\left\langle	\ds \frac{{\rmd} e_{K} }{{\rmd} T} \right\rangle= 0\,,\label{eq251a}
\end{eqnarray}
which means that $\left\langle e_{K} \right\rangle ={\rm const}$.

The differential equation \eref{eq202e} for osculating element $\omega$ after substituting for $S_K$ and $T_K$ from equations \eref{eq246}, \eref{eq247} reads
\begin{eqnarray}
 \frac{{\rmd}\omega_K}{{\rmd} T} = \ds \frac{\kappa n a^2 \left(1-e^2\right)^{\frac{1}{2}}}{e}\left(\ds \frac{9}{8}-\ds \frac{R}{a}\right) \cos f  \,.\label{eq252}
\end{eqnarray}
The average of each term of equation \eref{eq252} over a time period $P_b$ leads to
\begin{eqnarray}
	\left\langle\ds \frac{{\rmd}\omega_K}{{\rmd} T}\right\rangle = \ds \frac{3}{8} \kappa na \sqrt{ap}\,.\label{eq252a}
\end{eqnarray}

The differential equation \eref{eq202f} for the mean anomaly at the epoch $\mathfrak{M}$ after substituting for $S_K$ and $T_K$ from equations \eref{eq246}, \eref{eq247} gives
\begin{eqnarray}
		\ds \frac{{\rmd} \mathfrak{M}_{K}}{{\rmd} T} = - \left(1-e^2\right)^{\frac{1}{2}} \ds \frac{{\rmd}\omega_K}{{\rmd} T} + {2\kappa n a R}  \left(\ds \frac{9}{8}-\ds \frac{R}{a}\right) \,.\label{eq254}
\end{eqnarray}
The average of each term of equation \eref{eq254} over a time period $P_b$ result in
\begin{eqnarray}
	\left\langle \frac{{\rmd} \mathfrak{M}_{K}}{{\rmd} T}\right\rangle =  - \ds \frac{1}{8} \kappa n a^2\left(1+12e^2\right)\,.\label{eq254c}
\end{eqnarray}
This completes the calculations of the perturbations of the osculating elements.
\section{Discussion}\label{scCon}
We have used the exact solution of the Einstein equations found by McVittie \cite{mcvittie1933} in order to explore the time evolution of orbital elements of a test particle orbiting a massive spherically symmetric body (black hole) with mass $m$ caused by the Hubble expansion and cosmological curvature of space. The original McVittie metric is given in the global coordinates which are comoving with the Hubble observers and are not useful for interpretation of local physical experiments. To make our calculations physically meaningful, we have transformed the global coordinates $(t,r)$ to the local coordinates~$(T,R)$ of a Hubble observer associated with the central massive body $m$. McVittie's metric in the local coordinates was employed to derive the equation of motion of a test particle moving around the central mass $m$. We have found that the exact expressions for the gravitational perturbing forces are too complicated in the local coordinates and cannot be integrated analytically. In order to make the problem tractable we have expanded the perturbing forces with respect to small parameters: $HR/c\ll1$, ${m}/{R}\ll1$, $m/\mathfrak{a}\ll1$ up to the terms of the second-order and used them to determine the perturbation of the osculating elements with respect to time. We study the precession of the orbit of a test particle moving around the central body $m$ due to the cosmological expansion in the domain of $a \gg m$. 

The average size and shape of the elliptical orbit of the test particle, expressed in the local coordinates $(T,R)$, remain unaffected by cosmological expansion and spatial curvature to the order of $H_0^2$. However, they start evolving very slowly with time at higher-order terms of the order $H_0^3$. The argument of pericenter $\omega$ and the mean anomaly at the epoch $\mathfrak{M}$ change due to Hubble expansion and spatial curvature when terms of the order $H_0^2$ are considered. The inclination $i$ and the longitude of the ascending node $\Omega$ remain constant for any order of $H_0^2$ because of the conservation of the angular momentum of the test particle. Terms of the order $H_0$ are not present in the perturbation equations of the osculating elements due to Einstein’s equivalence principle.

The total rate of change of the argument of pericenter $\omega$ of a test particle with respect to time around the central body $m$ embedded in the cosmological manifold is given by $\left\langle\dot{\omega}\right\rangle = \left\langle\dot{\omega}_{pN} \right\rangle + \left\langle\dot {\omega}_H\right\rangle + \left\langle\dot{\omega}_K\right\rangle$, and can be calculated by using equations \eref{eq219b2}, \eref{eq239ub1}, and \eref{eq252a}. It yields
\begin{eqnarray}
	\hspace{-2.3cm}\left\langle \dot{\omega}\right\rangle =  \ds \frac{3 m n }{a\left(1- e^2\right)}\left[1-\ds \frac{\left(1-e^2\right)^{\frac{3}{2}}}{2n ^2}H^2\right] -\ds \frac{3 \left(1-e^2\right)^{\frac{1}{2}}}{2n}\mathfrak{q}H^2 \left(1-\ds \frac{4m}{3a}\right)+\ds \frac{3}{8} \kappa na \sqrt{ap}\,,\nonumber\\ \label{eq255}
\end{eqnarray}
where the first term in the right-hand side is the post-Newtonian precession, the terms proportional to $H_0^2$ are due to the cosmological expansion and the last term in the right-hand side is due to the spatial curvature of the universe. From equation \eref{eq255}, we conclude that when $\mathfrak{q}$ is negative, that is the universe is accelerating (expanding at an increasing rate), then the precession of the pericenter due to the cosmological expansion is positive. This moves the pericenter in the same direction to its post-Newtonian precession. Conversely, when $\mathfrak{q}$ is positive, that is the universe is decelerating (expanding at a decreasing rate), then the precession of the pericenter is negative. This results in the motion of the pericenter in the opposite direction to its post-Newtonian precession.

However, when $\mathfrak{q} = 0$, that is the universe is expanding at a constant rate ($\dot{\mathfrak{a}}={\rm constant}$), then the total effect from the $\mathfrak{q}$-term on the right hand side of equation~\eref{eq255} is zero which causes no precession of the pericenter. However, the second term on the right hand side of equation \eref{eq255} gives a negative value due to the Hubble expansion for the precession of the pericenter. This causes the pericenter to move in the reverse direction compared to its post-Newtonian precession. We compare the results of our calculations with the results of the other researchers \cite{2001_R, bolen2001, Mashhoon_2003, carrera2010, Arakida_2013}.

The results obtained by Kerr \cite{Mashhoon_2003} and Arakida \cite{Arakida_2013} on the perturbation of the argument of pericenter $\omega$ due to cosmological expansion, specifically equation $\left(38\right)$ in Kerr’s paper and equation $\left(11\right)$ in Arakida’s paper, are a particular case of our general equation \eref{eq255}. Indeed, their results can be derived from equation \eref{eq255} in our paper by setting $\mathfrak{q} = -1$, $k = 0$, and taking the limit $\left(m/a\right) \rightarrow 0$. The equation for the precession of the pericenter $ 3 \pi H^2 a^3/(GM) (1-e^2)^3$ for a nearly circular $(e \ll 1)$ approximation, obtained by Rindler \cite{2001_R}, Arakida \cite{Arakida_2013}, and Bolen \etal. \cite{bolen2001}, is incomplete. The incompleteness in Rindler’s and Arakida’s work is because they took only the first-order terms in $e$ to derive this equation from the Schwarzschild-de Sitter metric. On the other hand, the deficiency in the work by Bolen \textit{et al.} \cite{bolen2001} is due to their consideration of only the radial equation obtained from McVittie’s metric with $\dot{H} =0$, $k=0$ to derive their equation of motion for a test particle.

Antoniou and Perivolaropoulos \cite{Antoniou_2016arXiv} use the McVittie metric in comoving coordinates $(t, r, \theta, \phi)$ to describe the geodesics of a Schwarzschild spacetime embedded in an isotropic, expanding universe governed by phantom dark energy. The authors used the exact solution of the geodesic equations to calculate the time of orbital decay of binary systems, comparing it to the solution of the same problem in the Newtonian approximation. They found that the decay occurs earlier than predicted by the Newtonian approximation, but this difference is negligible for existing cosmological systems, such as galaxies or clusters of galaxies.

We make use of the equation \eref{eq255} in order to evaluate the numerical values for different types of the time rate of change of the argument of pericenter for few celestial bodies. The group of stars in close orbit around the supermassive black hole at the Milky Way galactic center are known as S stars. The black hole is named Sagittarius (Sgr) A*. We work on two S stars: S$2$, S$62$ and a binary star Sirius. The numerical values of different parameters for these celestial bodies and the evaluated values for different types of the time rate of change of the argument of pericenter by making use of the equation \eref{eq255} are given in \tref{table2}. The numerical values of Hubble's constant $H_0$, deceleration parameter $\mathfrak{q}_0$, and the accelerating expansion $\dot{H}_0$ used in these calculations are $2.3 \times 10^{-18}\; {\rm s}^{-1}$, $-0.55$, and $-2.3805 \times 10^{-36}\;{\rm s}^{-2}$ respectively.
\setlength{\tabcolsep}{2pt}
\renewcommand{\arraystretch}{2}
\begin{table} [ht]
\centering
\caption{\textbf{Precession rates due to the post-Newtonian approximation, the cosmological expansion and the spatial curvature of the universe for S$2$, S$62$, and  Sirius.}}
	\label{table2}
{\footnotesize{\begin{tabular}{c| c| c| c|c|c|c} 
\bhline
S stars  & Semi-major Axis & Eccentricity & Time Period   & $\left\langle\dot{\omega}_{pN} \right\rangle$ & $\left\langle\dot{\omega}_{H} \right\rangle$ &  $\left\langle\dot{\omega}_K\right\rangle$ for $k=+1$, \\[-1.3em]
and	Sirius & $a$~~(in meter) & $e$ & $P_b$ (in year)& {${\rm \mu a s}$}/{\rm yr} & {${\rm \mu a s}$}/{\rm yr} & {${\rm \mu a s}$}/{\rm yr} \cr
\hline
S$2$ & $1.45 \times 10^{14}$ & $0.88$ 
& $16.1$ & $4.78 \times 10^{7}$ & $1.0 \times 10^{-9}$ & $2.97 \times 10^{-18}$\cr
\hline 
S$62$ & $1.1 \times 10^{14}$ & $0.9$ 
& $9.9$ & $4.69 \times 10^{8}$ & $3.0 \times 10^{-10}$ & $1.31 \times 10^{-18}$\cr
\hline
Sirius & $2.95 \times 10^{12}$ & 0.59142 
& 50.1  & 189.27 & $5.70 \times 10^{-9}$ & $6.80 \times 10^{-22}$\cr
\bhline 
\end{tabular}}}
\end{table}

Carrera and Giulini \cite{carrera2010} reviewed the influence of global cosmological expansion on local systems. They explored how such influences can affect orbital motions and configurations of compact objects like black holes. Specifically, they investigated the dynamics of particles in McVittie’s spacetime. By substituting $R=a={\rm const}$ in the radial part of their geodesic equation, they derived a specific condition under which a test particle in McVittie’s spacetime (for a flat universe, $k=0$) exhibits non-expanding circular orbits. In terms of the
dimensionless quantities $h(t):= a H(t)$, $\ell:=L/a$ ($L$ -- the angular momentum of the test particle), and $\eta:=m/a$, the Carrera-Giulini's condition is
\begin{eqnarray}
	\ds a \dot{h} = \ds \frac{\left(1-2\eta -h^2\right)\left[\eta\left(1+3\ell^2\right)-\ell^2-h^2\right]}{\left(1+\ell^2\right)\sqrt{1-2\eta}}\,.\label{eq254c2}
\end{eqnarray}
In the case of a non-expanding circular orbit, $e=0$, $a={\rm const}.$, and $\ell=\sqrt{m/a}$, which brings equation \eref{eq254c2} to
\begin{eqnarray}
	\dot{H}= - H^2 \left( 1 -\ds \frac{2m}{a}\right) + \mathcal{O} \left(\ds \frac{m}{a}\right)^2\,.\label{eq254c3}
\end{eqnarray}
We can verify the consistency of Carrera-Giulini's condition \eref{eq254c3} within the framework of our formalism. To this end, we consider our equation \eref{eq231b} applied to the case of a non-expanding nearly-circular orbit, ${\rmd a_{H}}/{{\rmd} T}=0$. This constraint can be satisfied if, and only if, the terms on the right-hand side of equation \eref{eq231b} vanish identically. This yields: 
\begin{eqnarray} \mathfrak{q}= -\ds \frac{2m}{a} + \mathcal{O} \left(\ds \frac{m}{a}\right)^2\,,\label{eq254c4} 
\end{eqnarray} 
where $\mathfrak{q}$ has been defined in equation \eref{eq228d1}. Comparing \eref{eq254c4} with \eref{eq254c3}, we conclude that Carrera-Giulini's condition \eref{eq254c2} for non-expanding nearly-circular orbits is fully consistent with the results of our formalism.
Equation \eref{eq254c3}, considered as a differential equation for the Hubble parameter $H$, can be easily integrated with respect to time $T$, yielding: 
\begin{eqnarray} 
H T = \left(1 -\ds \frac{2m}{a}\right)^{-1}\,.\label{eq254c6} 
\end{eqnarray} 
Equation \eref{eq254c6} represents a constraint on the time dependence of the Hubble parameter to maintain the circular orbit during the universe's expansion. On the other hand, the time dependence of the Hubble parameter derived by solving the Friedmann equations is: \begin{eqnarray} H T= \ds \frac{2}{3(1+w) }\,,\label{eq254c7} 
\end{eqnarray} 
where $-1< w\le 1$ is a constant parameter characterizing the equation of state \cite{Liddle_2015}. Comparing equations \eref{eq254c6} and \eref{eq254c7}, we find that a particle in the expanding universe can have a circular orbit if, and only if, it has the orbit's radius $a$ satisfying the following condition: 
\begin{eqnarray} 1+3w = -\frac{4m}{a}\,.\label{eq254c8} 
\end{eqnarray} 
This constraint, however, can be valid only for $a\ge 0$, which corresponds to the universe with the equation of state parameter limited to the values $-1\le w\le -1/3$. This implies that the circular orbits of particles orbiting a black hole may exist only in the early universe governed by some form of dark energy, such as the cosmological constant or quintessence \cite{weinberg_2008}.

We proceed to discuss the perturbation in  mean anomaly at the epoch $\mathfrak{M}$ which gives the change in the frequency of the mean orbital motion $n$. We make a power series expansion of mean anomaly at the epoch $\mathfrak{M}$ around a point of time $T=T_0$, where $T_0$ is the epoch, as
\begin{eqnarray}
	\mathfrak{M} = \mathfrak{M}_0 + \dot{\mathfrak{M}}_0 (T-T_0) + \mathcal{O} \left[\left(T-T_0\right)^2\right]\,,\label{eq242lb1}
\end{eqnarray}
where $\mathfrak{M}_0$ is the mean anomaly at the epoch at $T=T_0$ and $\dot{\mathfrak{M}}_0=\ds \frac{{\rmd} \mathfrak{M}}{{\rmd} T}\Bigg|_0$. By comparing equation \eref{eq242lb1} with equation \eref{eq200}, we see that the second term of equation \eref{eq242lb1} couples with the second term in equation \eref{eq200}. The frequency of the mean orbital motion $n$ can be written as 
\begin{eqnarray}
	n = {n_0} + \langle{\dot{\mathfrak{M}}}\rangle \,,\label{eq242lb1a}
\end{eqnarray}
where
\begin{eqnarray}
\langle{\dot{\mathfrak{M}}}\rangle = \langle{\dot{\mathfrak{M}}_{pN}}\rangle + \langle{\dot{\mathfrak{M}}_H}\rangle + \langle{\dot{\mathfrak{M}}_K}\rangle\,. \label{eq242lb2}
\end{eqnarray}
We substitute the values of $\langle{\dot{\mathfrak{M}}_{pN}}\rangle$, $\langle{\dot{\mathfrak{M}}_H}\rangle$, and  $\langle{\dot{\mathfrak{M}}_K}\rangle$ from equations \eref{eq219b1b}, \eref{eq242lb}, \eref{eq254c} in equation \eref{eq242lb2}. It yields 
\begin{eqnarray}
\langle{\dot{\mathfrak{M}}}\rangle  =  & \ds \frac{3 m n}{2a } \left\{4 - \ds \frac{6 }{\left(1-e^2\right)^{\frac{1}{2}}} +  \ds \frac{\left(1-e^2\right)}{n^2}H^2\left(1-\ds \frac{8}{3}\ds \frac{1+e^2}{1-e^2}\right)\right\}
 \nonumber\\
& + \ds \frac{3 \left(1-e^2\right)}{2n} \mathfrak{q} H^2 \left[1-\ds \frac{4m}{3a} + \ds \frac{4 }{3\left(1-e^2\right)}\left(1+\ds \frac{3}{2}e^2+\ds \frac{2m}{a}\right)\right] \nonumber\\
&
 - \ds \frac{1}{8} \kappa n a^2\left(1+12e^2\right). \label{eq242lb3}
\end{eqnarray}
The first two terms on the right-hand side of \eref{eq242lb3} explain the change in the mean orbital motion  $n$ due to the post-Newtonian terms. The terms of order $H^2$ change the mean orbital motion $n$ due to the cosmological expansion.

We observe that when $\mathfrak{q} < 0$, the $\mathfrak{q}$-term on the right-hand side of \eref{eq242lb3} is negative, which decreases the mean orbital motion $n$.
Conversely, when $\mathfrak{q}>0$, the $\mathfrak{q}$-term on the right-hand side of \eref{eq242lb3} is positive, which increases the mean orbital motion $n$. 
However, when $\mathfrak{q} = 0$, that is, the universe is expanding at a constant rate ($\dot{\mathfrak{a}}={\rm constant}$), the total effect from the $\mathfrak{q}$-term on the right hand side of equation \eref{eq242lb3} is zero which causes no change in the mean orbital motion  $n$.

The last term on the right-hand side of equation \eref{eq242lb3} explains the change in mean orbital motion due to the spatial curvature of the universe. The value for mean orbital motion $n$ increases for $k=-1$, decreases for $k=+1$, and remains same for $k=0$.

\ack
We appreciate the insightful comments from the anonymous referees, which helped us enhance the presentation of our paper and the discussion on gravitational physics. One of us (V.J.) is particularly grateful to Dr. Brett Bolen for his valuable conversations and clarifications regarding the problem discussed in this manuscript.

\appendix
\section{McVittie's metric in the local coordinates for flat universe $k=0$ as Kottler's metric}\label{kottler}
In this appendix we prove that Kottler's metric (Schwarzschild-de Sitter metric) \cite{doi:10.1080/14786440508564528} is McVittie's metric in the local coordinates for flat universe $k=0$. We substitute $f_1$, $g_1$ from  equations \eref{eq177i}, \eref{eq182c}, $G_2=H R$ in equations \eref{eq185a} and \eref{eq185} to obtain ${G}_{T T}(T,R)$ component and ${G}_{RR}(T,R)$ component of Kottler's metric. We make use of equations~\eref{eq187b}, \eref{eq187a} with the condition imposed: spatial curvature $k=0$ and $H$ is constant, to obtain
\begin{eqnarray}
{G}_{TT}(T,R) = -\left(1-\ds \frac{2 m}{R}- H^2R^2\right)\,,\label{eq204o1}\\
{G}_{R R}(T,R)	 = \left(1-\ds \frac{2 m}{R}- H^2R^2\right)^{-1}\,.\label{eq204o2}
\end{eqnarray} 
As the Kottler metric is for flat universe $k=0$, we should use the value of $\chi(T,R)$ for $k=0$. We proceed to calculate $\chi(T,R)$ for $k=0$ by making use of equations  \eref{eq158dd}, \eref{eq177i}, \eref{eq182c}, \eref{eq187b}, \eref{eq187a} in equation \eref{eq182b1}
\begin{eqnarray}\label{eq204o3}
	\chi(T,R) = \int \ds \frac{F_1G_2}{G_1^2 - G_2^2} {\rmd} R + C(T)\,,
\end{eqnarray}	
where $C(T)$ is the integration constant. $F_1$ and $G_1$ are expressed below
\begin{eqnarray}\label{eq204o4}
	F_1 = \ds \frac{1}{x} - \ds \frac{K (1+x)^3}{4 (1-x)\,x^3}+\mathcal{O} \left(K^2\right)\,,
\end{eqnarray}
\begin{eqnarray}\label{eq204o5}
	G_1= x+\ds \frac{K (1+x)^3 (1-5x)}{4 (1-x)^2 x}+ \mathcal{O}\left(K^2\right)\,,
\end{eqnarray}	
where $K \equiv \ds \frac{\kappa m^2}{16}$, $\kappa \equiv {k}/{\mathfrak{a}^2(T)}$, and we have dropped out the residual terms of the order $K^2$ because they are negligibly small.
We substitute for $F_1$ from equation \eref{eq204o4} and $G_1$ from equation \eref{eq204o5} in equation \eref{eq204o3}. We get
\begin{eqnarray}\label{eq204o6}	
	\chi(T,R) = 8 H m^2 \int \ds \frac{{\rmd} x}{\left[x^2 \left(1-x^2\right)^2-4 H^2 m^2 \right]\left(1-x^2\right)} \,,
\end{eqnarray}
where $R ={2 m}/{(1 - x^2)}$, ${\rmd} R =  {4 m x}\; {\rmd} x/{(1 - x^2)^2}$ and the integration constant $C(T)$ is absorbed into the definition of new time coordinate $T$.

In this way the metric coefficients of the Kottler metric that are equation \eref{eq204o1} and equation \eref{eq204o2} are derived from McVittie's metric in the local coordinates for flat universe $k=0$.
\section{The Hansen coefficients}\label{scap}
Using equation \eref{eq204b} we calculate the following Hansen coefficients:
\begin{eqnarray}
	\left\langle {\cos f} \right\rangle = 
	X_0^{0,1}(e) =  -e\,,\label{eq204k}\\
	\left\langle {\cos 2f} \right\rangle = 
	X_0^{0,2}(e) =  \ds \frac{-1+2e^{2}+(1-e^{2})^{1/2}}{1+(1-e^{2})^{1/2}} \,,\label{eq204l}\\
	\left\langle {\ds \frac{R}{a} \cos f} \right\rangle = 
	X_0^{1,1}(e) =  -\ds \frac{3e}{2}\,,\label{eq204g}\\
	\left\langle {\ds \frac{R}{a} \cos 2f} \right\rangle = 
	X_0^{1,2}(e) =  \ds \frac{3e^{2}}{2}\,,\label{eq204h}\\
	 \left\langle {\ds \frac{R}{a}} \right\rangle = 
	X_0^{1,0}(e) =  1+\ds \frac{e^{2}}{2}\,,\label{eq204c}\\
	 \left\langle \ds \frac{R^2}{a^2} \right\rangle = 
	X_0^{2,0}(e) = 1 + \ds \frac{3e^{2}}{2} ,\label{eq204d}\\
	\left\langle {\ds \frac{a}{R} \cos f} \right\rangle = 
	X_0^{-1,1}(e) =  -\ds \frac{e}{1+(1-e^{2})^{1/2}}\,,\label{eq204i}\\
	\left\langle{ \ds \frac{a}{R}}\right\rangle = 	X_0^{-1,0}(e) =1\,,\label{eq204v1}\\
	\left\langle{ \ds \frac{a^2}{R^2}}\right\rangle = 	X_0^{-2,0}(e) =\ds \frac{1}{\left(1-e^2\right)^{1/2}}\,,\label{eq204v}\\
	\left\langle{\ds \frac{a^3}{R^3}}\right\rangle = 	X_0^{-3,0}(e)=\ds \frac{1}{\left(1-e^2\right)^{3/2}} \,,\label{eq204u}\\
	\left\langle{\ds \frac{a^4}{R^4}}\right\rangle =X_0^{-4,0}(e)=\ds \frac{1+\ds \frac{e^2}{2}}{ \left(1-e^2\right)^{5/2}}\,,\label{eq204t}\\
	\left\langle{\ds \frac{a^5}{R^5}}\right\rangle =X_0^{-5,0}(e)=\ds \frac{1+\ds \frac{3e^2}{2} }{ \left(1-e^2\right)^{7/2}} \,.\label{eq204s}
\end{eqnarray}
\section{Exact Asymptotic Solution for Elliptical Orbits}
Due to the fully nonlinear nature of the Einstein field equations, it is not unreasonable to expect that the explicit orbit of a test body for small values of the expansion parameters may differ significantly from the one derived by applying the averaging technique. This appendix provides a theoretical argument demonstrating that the error between the solution of the equation of motion presented in the main text and the actual solution is small.

Let us assume that the orbit of a particle is elliptic at the initial instant of time, $a=a_0$, $e=e_0$, and consider the evolution of this orbit in a one-component universe with the Hubble parameter described by the equation $H=\frac{2}{3} \frac{1}{\left(1+w\right)T}$, $\left(w\neq -1\right)$. We consider equations \eref{eq231b} and \eref{eq231bb} for the evolution of the semi-major axis and eccentricity, retaining only the principal terms by neglecting all terms proportional to the mass $m$ of the central black hole: 
\begin{eqnarray} 
\frac{{\rmd a}}{{\rmd} T} = -\ds \frac{2e}{n\left(1-e^2\right)^{\frac{1}{2}}}H^2\mathfrak{q} R \sin f\,,\label{eq231ba} 
\end{eqnarray} 
\begin{eqnarray} 
\frac{{\rmd} e }{{\rmd} T} = -\ds \frac{\left(1-e^2\right)^{\frac{1}{2}}}{n a}H^2 \mathfrak{q} R \sin f\,.\label{eqaa} 
\end{eqnarray} 
The first integral of these two equations is 
\begin{eqnarray}
a \left(1-e^2\right) =p= {\rm const.}\,,\label{eqb} 
\end{eqnarray}
which is the focal parameter of the orbital motion of the particle. From the law of conservation of the focal parameter, it follows that if $e$ increases, $a$ increases and vice-versa. Therefore, it is sufficient to find the time evolution of the eccentricity by integrating equation \eref{eqaa}. Expressing time $T$ in terms of the eccentric anomaly $E$ by using equations \eref{eq197a} and \eref{eq200}, and replacing $R \sin f = a \sqrt{1-e^2} \sin E$ in equation \eref{eqaa}, we get 
\begin{eqnarray} 
\frac{1}{1-e^2}\frac{{\rmd} e}{{\rmd} E} = - \ds \frac{2}{9}\ds \frac{1+3w}{\left(1+w\right)^2} \ds \frac{1-e \cos E}{\left(E-e\sin E\right)^2} \sin E\,.\label{eqac} 
\end{eqnarray} 
We consider the asymptotic solution of this equation for very large values of time $T$, that is, for $E \to \infty$. In this case, we can drop the terms with eccentricity $e$ on the right-hand side of equation \eref{eqac}, which makes it analytically integrable:
\begin{eqnarray} 
e = \tanh\left[c_0- \ds \frac{2}{9}\ds \frac{1+3w}{\left(1+w\right)^2} \left(\ds \frac{\sin E}{E}-\int_{E}^{\infty} \ds \frac{\cos x}{x} {\rm d}x \right)\right]\,,\label{eqad} \end{eqnarray} 
where $c_0$ is a constant of integration defining the value of eccentricity $e_0$ at the initial instant of time. Asymptotic expansion of the integral on the right-hand side of this equation with respect to $E\to\infty$ yields: 
\begin{eqnarray} 
e = \tanh\left[c_0+\ds \frac{2}{9} \ds \frac{1+3w}{\left(1+w\right)^2}\ds \frac{\cos E}{E^2} \right]\,.\label{eqae} 
\end{eqnarray}
This result shows that the eccentricity of the orbit oscillates around its constant initial value $e_0$ with the orbital frequency and asymptotically approaches it as time grows, $E \to \infty$. The semi-major axis experiences the same type of behavior as a consequence of the first integral of motion \eref{eqb}. It oscillates around its initial value $a_0$, asymptotically approaching it as the universe expands. These results are fully consistent with those derived in \sref{scH} by applying the averaging technique and make those results robust.

We can also see from equation \eref{eqae} that ideal circular orbits are allowed only if the parameter of the equation of state of the universe is $w=-\frac{1}{3}$, which also agrees with the constraint \eref{eq254c8}. Consideration of the asymptotic behavior of elliptical orbits in the case of a universe governed by the cosmological constant $(w=-1, H={\rm const.}, q=-1)$ can be done in a similar way. In this scenario, equations \eref{eq231ba} and \eref{eqaa} can also be solved by applying asymptotic expansion with respect to the eccentric anomaly $E$, and the conclusion about the asymptotic behavior of the orbital elements $a$ and $e$ is the same -- they oscillate around the initial value, asymptotically approaching it.

\section*{References}
\bibliographystyle{iopart-num}
\bibliography{CQG}
\end{document}